\documentclass{jfmc}

\usepackage{graphicx}
\usepackage[normalem]{ulem}
\usepackage{newtxtext}
\usepackage{newtxmath}
\usepackage{natbib}
\usepackage{hyperref}
\hypersetup{
    colorlinks = true,
    urlcolor   = blue,
    citecolor  = black,
}

\newcommand{\RomanNumeralCaps}[1]

\title{Asymptotic solutions for self-similarly expanding fault slip induced by fluid injection at constant rate}

\author{Robert C. Viesca\aff{1}
  \corresp{\email{robert.viesca@tufts.edu}}}

\affiliation{\aff{1}Department of Civil and Environmental Engineering, Tufts University, Medford, MA 02155, USA}

\begin{document}
\maketitle

\begin{abstract}
We examine the circular, self-similar expansion of frictional rupture due to fluid injected at a constant rate. Fluid migrates within a thin permeable layer parallel to and containing the fault plane. When the Poisson ratio $\nu=0$, self-similarity of the fluid pressure implies fault slip also evolves in an axisymmetric, self-similar manner, reducing the three-dimensional problem for the evolution of fault slip to a single self-similar dimension. The rupture radius grows as $\lambda \sqrt{4\alpha_{hy} t}$, where $t$ is time since the start of injection and $\alpha_{hy}$ is the hydraulic diffusivity of the pore fluid pressure. The prefactor $\lambda$ is determined by a single parameter, $T$, which depends on the pre-injection stress state and injection conditions. The prefactor has the range $0<\lambda<\infty$, the lower and upper limits of which correspond to marginal pressurization of the fault and critically stressed conditions, in which the fault-resolved shear stress is close to the pre-injection fault strength. In both limits, we derive solutions for slip by perturbation expansion, to arbitrary order. In the marginally pressurized limit ($\lambda\rightarrow 0$), the perturbation is regular and the series expansion is convergent. For the critically stressed limit ($\lambda\rightarrow \infty$), the perturbation is singular, contains a boundary layer and an outer solution, and the series is divergent. In this case, we provide a composite solution with uniform convergence over the entire rupture using a matched asymptotic expansion. We provide error estimates of the asymptotic expansions in both limits, and demonstrate optimal truncation of the singular perturbation in the critically stressed limit.
\end{abstract}

\section{Introduction}
\label{sec:headings}
Boundary layers and matched asymptotic perturbations are implicit in many problems in fracture mechanics, but details are often subsumed within the so-called stress intensity factor. This factor provides the leading-order matching condition between an outer solution on the scale of the entire fracture and an inner solution representing a physical---often weakening---process in the vicinity of the crack tip. When the small-scale-yielding condition for the separation of scales between the entire fracture and this near-tip cohesive zone is met, the details of the breakdown process are often neglected, yet assumed to be universal for a given material such that a critical value of the stress intensity factor for a propagating crack is assumed to be a material constant. As a consequence, a cohesive-zone boundary layer solution near the tip is seldom explicitly sought and even more rarely matched with an outer solution to provide a composite solution of crack-face displacement along the entire crack. In practice this step is often unnecessary, but may be a missed opportunity to demonstrate a more fundamental understanding of models involving fracture and to exhibit perturbation methods as applied to fracture. 

\

There nonetheless has been a substantial series of works that have pursued singular perturbations involving boundary layers in classical fracture problems and in those coupling fracture with other physical processes.  
In a series of works, \emph{Edmunds and Willis} [1976a,b, 1977] examined higher-order matching of the far-field with the near-field of a crack-tip cohesive zone. Among other results, the authors demonstrated a correction to the crack length appearing in standard expressions for the stress-intensity factor, the corrections being equivalent to those of second-order to the energy release rate.
For the specific geometry of a long, thin crack running parallel to a free surface, as may occur in spalling, several workers examined an inner solution corresponding to a semi-infinite crack as well as its matching with an outer solution corresponding to the thin ligament deforming as a simple Euler-Bernoulli beam [\emph{Zlatin and Krapkov}, 1986; \emph{Thouless et al.}, 1987; \emph{Dyskin et al.}, 2000]. \emph{Lister et al.} [2019] performed similar analysis considering the more elaborate case where the crack is being driven by the migration of a viscous fluid. 
Such hydraulic fracture in its own right has been a source of novel problems in the asymptotic, boundary-layer analysis of fractures. Early work included self-similar solutions and asymptotic examination of near-tip viscous flow [\emph{Spence and Sharp}, 1985; \emph{Lister, 1990}; \emph{Desroches et al.}, 1994; \emph{Garagash and Detournay}, 2000]. This was followed by matched-asymptotic expansions of self-similarly expanding hydraulic fractures in the so-called viscosity dominated regime, in which, to leading order, effects of fracture toughness are confined to a near-tip boundary layer [\emph{Garagash and Detournay}, 2005], as well as regular asymptotic expansions in the so-called large-toughness regime [\emph{Savitski and Detournay}, 2002; \emph{Garagash}, 2006]. 
Boundary-layer solutions also appear in problems of the fracture of poroelastic media. \emph{Atkinson and Craster} [1994] examined the sudden appearance of a finite, stress-free crack in a porous medium. This problem has an early-time solution that can be considered to be the composition of an ``undrained" outer-solution over most of the fracture, in which fluid flow is neglected, as well as an inner solution valid near the crack tips, the surroundings of which are in an effectively ``drained" state and for which fluid flow is sufficiently quick as to instantaneously establish a constant, uniform fluid pressure. An earlier series of works also considered the steady propagation of a semi-infinite fracture in a poroelastic medium for a variety of permeability conditions along the fault, and in which boundary layers appear in the vicinity of the crack tip [\emph{Rice and Simons}, 1976; \emph{Simons}, 1977; \emph{Rudnicki}, 1991; \emph{Rudnicki and Koutsibelas}, 1991; \emph{Craster and Atkinson}, 1992, 1994].
 In all of the above-mentioned works, perturbations are carried out to within two terms, with the exception of the work of \emph{Garagash} [2006] who provides results of numerical solutions for finitely higher-order expansions.

\

Here we look to a model problem that provides the opportunity to examine both regular and singular perturbation methods to arbitrary order and to demonstrate optimal truncation of the latter. 
The model problem consists of a singular integral equation for a circular crack in a three-dimensional medium the growth of which is driven by fluid injection at constant volume rate into a narrow permeable region surrounding a planar geologic fault, such that fluids permeate away from the injection site and remain within the fault plane. 
The injection elevates pore fluid pressure along the fault and reduces the effective normal stress carried by solid contacts, and hence the frictional strength. 
We will investigate an elementary scenario in which the initial the stress state is uniform and the fault's frictional strength follows a Coulomb friction law.
Injection of fluid into the fault plane---and its subsequent migration along that plane---elevates fluid pressure there, which drives the loss of strength and permits the nucleation and growth of a crack.
Because the functional form of the forcing is known \emph{a priori}, we can directly express a self-similar solution for slip and crack radius in integral form  using a solution of \emph{Sneddon} [1951].
This solution cannot be calculated directly and must be evaluated numerically. 
However, asymptotic solutions can be found using perturbation techniques.
The problem has one parameter with two end-member limits: in one limit we derive a regular perturbation expansion, while in the other we derive a singular perturbation expansion accounting for the existence of a boundary layer.
The boundary layer results from a separation of the length scale characteristic of the forcing driving the crack from the size of the crack itself.
In the limit in which the boundary layer exists, we will not need to explicitly match the asymptotic expansions to connect the so-called inner solution of the boundary layer with the outer solution away from the layer. Rather, the inner and outer solutions we calculate will inherently satisfy matching conditions, owing to the use of the inversion formula. 
Previously, we performed a comparable analysis for the self-similar growth of a frictional mode-II or -III fracture with a constant friction coefficient under fluid injection at constant pressure, providing the first two terms in the perturbation series solution in each end-member limit [\emph{Viesca} 2021]; leading-order asymptotics in these limits were also sought and solved for by Garagash and Germanovich [2012] via a mix of closed-form and numerical techniques for the case of a mode-II or -III fracture under the same injection conditions, except with slip-weakening friction, which precluded self-similarity but permitted dynamic rupture nucleation. 
Here we address the mixed-mode problem of a circular shear crack.

\

This problem was previously considered numerically by \emph{S\'aez et al.} [2022] who considered the more general free-boundary problem in which in which the crack boundary is not known \emph{a priori} and part of the solution.
When the Poisson ratio $\nu=0$, the crack boundary is expected to be circular and grow in a self-similar manner.
In this axisymmetric case, numerical solutions were provided spanning the range of the problem parameter.
In addition, leading-order asymptotic solutions in this circular-crack case were provided in the limit of two end-member values of the problem parameter: a complete leading-order solution was provided in one limit, and an outer solution provided in the other.
A boundary element method was employed to solve the free-boundary problem when $\nu>0$ and the numerical axisymmetric solutions, in addition to the asymptotic behavior, were used to verify boundary-element solutions in the $\nu=0$ limit. 
The solutions of \emph{S\'aez et al.} [2022] demonstrated that, for cases in which $\nu\neq 0$, crack growth also proceeds self-similarly and its boundary is nearly elliptical.
These solutions also demonstrated that the growth of the elliptical rupture area ($\nu\neq 0$) was, to within the accuracy of the solutions, equal to that of the circular crack ($\nu=0$) for a given value of the problem parameter.
This has led to a renewed interest in using a circular-crack approximation as a reduced-order model for related problems [\emph{S\'aez and Lecampion}, 2023; \emph{S\'aez and Lecampion}, 2024], as well as a benchmark. 
Here we return to this problem and provide the complete asymptotic expansions in both limits of the problem parameter, including the full inner and outer solutions in the limit for which singular perturbations occur and a boundary layer exists.

\

Independent of its perturbation aspects, this problem is of interest to understand the activation of geologic faults in the subsurface in response to transiently elevated pore fluid pressure.
The fundamental and industrial interest for such problems range from understanding the interplay between natural fluid migration and aseismic or seismic fault activity to monitoring of the artificial injection of fluids, such as to avoid the activation of faults to preserve the integriy of an imperable caprock formation, or to intentionally enhance the permeability of naturally occurring faults for the extraction of geothermal energy by inducing their slip.
In any context, non-numerical solutions for models of such problems are rare.
Such solutions provide fundamental scaling relations for the propagation of the rupture front and the accumulation of fault slip, in addition to its distribution, as they may vary with the strength of the injection and the pre-existing state of stress on the fault.
Comprehensive analytical solutions are typically only available when model complexity is low, as is the case for the elementary model examined here. 
Nonetheless, these solutions provide a basis to understand how increasing model complexity alters model behavior.
Areas of potential complexity in these models may include representations of the fault hydrology, friction, or geometry. 
For example, several prior works have examined the evolution of fault slip following fluid injection assuming friction laws alternative to the Coulomb law assumed here, such as slip-weakening or rate- and state-dependent friction [e.g., \emph{Garagash and Germanovich}, 2012; \emph{Viesca and Rice}, 2012; \emph{Dublanchet}, 2019; \emph{Im and Avouac}, 2024]. These have predominately examined ruptures within a two-dimensional medium, such as for in-plane or anti-plane deformation. 

\

The outline of this work is as follows. 
In section \ref{sec:pf}, we formulate the problem determining the fault slip distribution and rupture front location. We provide an integral representation of the solution for these quantities. We provide asymptotic expressions for the solution of the rupture front location as it depends on the pre-injection state of fault traction and strength and a diffusive lengthscale $\sqrt{\alpha t}$.
In section \ref{sec:cs}, we provide singular perturbation expansions of this solution in the so-called critically stressed limit of the problem parameter. In this limit, the solution for slip contains a boundary layer at the center of the crack, about the injection point, over distances comparable to the diffusive lengthscale. We provide the expansion for this inner solution to arbitrary order in the perturbation parameter. Over distances comparable to the crack radius but away from the diffusive boundary layer, the details of fluid diffusion are negligible, and we likewise provide the expansion for the so-called outer behavior of the solution to arbitrary order. This section demonstrates the asymptotic matching of the inner and outer solutions, calculates their overlap, and constructions a composite solution that uniformly approximates the solution over the entire rupture.
In section \ref{sec:mp}, we provide regular perturbation expansion of the solution for slip in the so-called marginally pressurized limit of the problem parameter.
In section \ref{sec:Md0}, we examine the accumulation of the peak slip and aseismic moment across all values of the problem parameter and provide asymptotic expansions for both quantities in the critically stressed and marginally pressurized limits. 
In section \ref{sec:ee} we provide error estimates of the singular and regular perturbation series in the two limits and demonstrate optimal truncation of the singular perturbation series. Finally, in section \ref{sec:dis} we discuss several issues concerning the presented solutions, including the choice of perturbation parameter, the importance of the solution to a problem for which $\nu=0$, and relevant applications.

\section{Problem formulation}
\label{sec:pf}

For a circular shear fracture of radius $R(t)$ undergoing relative slip $\delta(r,t)$ in a single direction, the shear traction on the crack faces $\tau(r,t)$, opposing the direction of slip, is given by [\emph{Salamon and Dundurs}, 1971, 1977; \emph{Viesca and Bhattacharya}, 2019]
\begin{equation}
\tau(r,t)=\tau_b+\frac{\mu}{2\pi}\int_0^{R(t)} \left(\frac{E[k(r/s)]}{s-r}+\frac{K[k(r/s)]}{r+s}\right)\frac{\partial \delta(s,t)}{\partial s}ds
\label{eq:td}
\end{equation}
where $k(u)=2\sqrt{u}/(1+u)$, and $E$ and $K$ are the complete elliptical integrals of the second and first kind. The background shear stress resolved on the plane of the fracture in the absence of slip, $\tau_b$ is here taken to be uniform. The relation (\ref{eq:td}) presumes that the Poisson ratio $\nu=0$. This expression is equivalent to that relating relative crack opening and crack-face tractions for a circular mode-I crack. In that case, this expression holds for any value of the Poisson ratio, provided that $\mu$ here is replaced by $\mu/(1-\nu)$. The inversion of this equation for the relative displacement $\delta$ was done in the context of mode-I equation [\emph{Sneddon, 1951}], which we adapt here for our particular mixed-mode II/III case with $\nu=0$
\begin{equation}
\delta(\rho,t)=\frac{4}{\pi}\frac{R(t)}{\mu}\int_\rho^1\frac{\eta}{\sqrt{\eta^2-\rho^2}}\int_0^1\frac{x\,\Delta \tau(x\eta R,t)}{\sqrt{1-x^2}}dx\,d\eta
\label{eq:dt}
\end{equation}
where 
\begin{equation}
\Delta\tau(r,t)=\tau_b-\tau(r,t)
\label{eq:dtau}
\end{equation}
and $\rho=r/R(t)$.

\

We may rearrange this inversion to isolate the contribution of a crack-tip singularity to the slip distribution. To do so, we first denote the innermost integral 
\begin{equation*}g(\eta)=\int_0^1\frac{x\,\Delta \tau(x\eta R,t)}{\sqrt{1-x^2}}dx\end{equation*}
where for brevity and without loss of generality, we have suppressed explicit mention of the time dependence of $g$.
The inversion (\ref{eq:dt}) reduces to
\begin{equation}
\delta(\rho,t)=\frac{4}{\pi}\frac{R(t)}{\mu}\int_\rho^1\frac{\eta}{\sqrt{\eta^2-\rho^2}}g(\eta)\,d\eta
\label{eq:dtf}
\end{equation}
Integration of (\ref{eq:dtf}) by parts yields
\begin{equation}
\frac{\delta(\rho,t)}{4 R(t)/(\pi \mu)}=g(1)\sqrt{1-\rho^2}-\int_\rho^1\sqrt{\eta^2-\rho^2}\frac{dg}{d\eta}d\eta
\label{eq:dtfbp}
\end{equation}
where we identify that
\begin{equation}
g(1)=\int_0^1 \frac{x \Delta\tau(x R,t)}{\sqrt{1-x^2}}dx
\label{eq:f1}
\end{equation}
is a stress intensity factor as it is a pre-factor for the relative-displacement field behind the crack tip, $\delta\sim g(1)\sqrt{1-r}$ corresponding to a singular stress field ahead of the crack tip $\tau\sim g(1)/\sqrt{r-1}$. This shear stress intensity factor is identically zero for problems in which the shear stress must remain bounded, without singularity, owing to the finite strength of the fault, the source for which is Coulomb friction.

\

We now look to consider a particular form of $\Delta\tau$ corresponding to the injection of fluid at a constant volumetric rate $Q$ into a thin fault zone whose Coulomb friction coefficient remains at a constant value, $f$. The strength of the fault is given by the product of $f$ and the effective normal stress $\sigma-p$,
\begin{equation}
\tau_s=f(\sigma -p(r,t))
\label{eq:ts}
\end{equation}
where $\sigma$ is the total normal stress, which is constant and uniform along the fault, $p$ is the local fluid pressure at a point on the fault, which is presumed to have an initial, uniform value $p_o$ plus a perturbation due to injection, given by the solution 
to the axisymmetric diffusion equation governing fluid pressure in the plane of the thin fault zone
\begin{equation*}\frac{\partial p}{\partial t}=\alpha_{hy}\frac{1}{r}\frac{\partial}{\partial r}\left(r\frac{\partial p}{\partial r} \right)\end{equation*}
where $\alpha_{hy}=k/(\eta_f \beta)$ is the hydraulic diffusivity $k$ is the fault-zone permeability, $\eta_f$ is the fluid viscosity and $\beta$ is the composite compressibility of the porous matrix and fluid. The corresponding solution of interest is that subject to the injection condition
\begin{equation*}-2\pi r k\frac{\partial p}{\partial r}=\frac{Q}{h}\end{equation*}
and is given by
\begin{equation}
p(r,t)=p_o+\Delta p E_1[r^2/(\alpha t)]
\label{eq:p}
\end{equation}
where $E_1$ is the exponential integral, $\alpha = 4\alpha_{hy}$, and $\Delta p = Q/(4\pi k h)$. The thinness of the fault zone is relevant when considering that the fault zone as a permeable, poroelastic conduit confined within an elastic medium, which may have a stiffness dissimilar to the fault zone. Here we consider the case where the fault-zone thickness $h$ is much smaller than the diffusive lengthscale $\sqrt{\alpha t}$. Under this condition, there is no change in the fault-normal stress over distances comparable to $\sqrt{\alpha t}$ due to localized, constrained swelling of the fault zone around the injection source. In this case fluid pressure is decoupled from volumetric straining of the fault zone and follows the autonomous diffusion equation [\emph{Marck et al.}, 2015 consider the injection scenario here; see also \emph{Jacquey and Viesca}, 2023, for similar results under plane-strain conditions; earlier works, such as \emph{Garagash and Germanovich} [2012], have implicitly operated in this limit]. 

\

Self-similarity of the pore-fluid pressure suggests self-similarity of the rupture front and slip profile. In this case, the rupture radius may be written as
\begin{equation}
R(t)=\lambda\sqrt{\alpha t}
\label{eq:R}
\end{equation}
where $\lambda$ is a pre-factor to be determined. The non-singular stress condition, $f(1)=0$, provides a condition to solve for the pre-factor $\lambda$ in terms of other problem parameters. This condition is imposed because the finite-frictional strength of the fault precludes a stress singularity. To impose this non-singular condition, we note that the frictional strength condition requires fault shear stress to equal fault shear strength, $\tau(r,t)=\tau_s(r,t)$, when and where sliding occurs. Combining this equation with (\ref{eq:dtau}), and (\ref{eq:f1}--\ref{eq:R}), and rearranging,
 \begin{equation}
T=\int_0^1 \frac{x\, E_1\!\left[(\lambda x)^2\right]}{\sqrt{1-x^2}}dx
\label{eq:fT0}
\end{equation}
where the dimensionless parameter
\begin{equation}
T=\frac{\sigma'}{\Delta p}\left(1-\frac{\tau_b}{\tau_p}\right)
\label{eq:T}
\end{equation}
is a measure of distance of the fault stress state from critically stressed conditions, $\tau_b\rightarrow \tau_p$, in which $T\rightarrow 0$ and the pre-injection strength of the fault is $\tau_p=f\sigma'$, and also where $\sigma'=\sigma-p_o$ is the pre-injection effective fault normal stress. When the initial stress $\tau_b$ is a finite distance away from $\tau_p$, but the injection-induced pressure coefficient $\Delta p$ is small compared with $\sigma'$, the fault is marginally pressurized and $T\rightarrow\infty$.

\

We may anticipate the behavior of $\lambda$ in these two end-member regimes: i.e., we expect that the rupture will progress quickly relatively to pore-fluid pressure diffusion ($\lambda \gg1$) when the fault is critically stressed  and relatively slowly ($\lambda\ll1$) when marginally pressurized. While a full solution to (\ref{eq:fT0}) was stated by \emph{S\'aez et al.} [2022] as an infinite (hypergeometric) series, we can derive concise expressions of the relation between $T$ and $\lambda$ in these limits by asymptotically expanding the integral (\ref{eq:fT0}).

\

For large values of $\lambda$, we apply the change of variable $u=x\lambda$
 \begin{equation}
T=\frac{1}{\lambda^2}\int_0^\lambda \frac{u\, E_1\!\left[u^2\right]}{\sqrt{1-(u/\lambda)^2}}du
\label{eq:fT}
\end{equation}
followed by the Taylor expansion
\begin{equation}
\frac{1}{\sqrt{1-(u/\lambda)^2}}=1+\frac{(u/\lambda)^2}{2}+...+\frac{1}{\sqrt{\pi}}\frac{[(n-1)/2]!}{(n/2)!}(u/\lambda)^n
\end{equation}
for even $n$, and
 \begin{equation}
T= \sum_{n=0,2,...} ^\infty \frac{1}{\lambda^{2+n}} \frac{1}{\sqrt{\pi}}\frac{[(n-1)/2]!}{(n/2)!} \int_0^\lambda u^{n+1}\, E_1\!\left(u^2\right)du
\label{eq:lcs1}
\end{equation}
Passing the limit of the integrals in (\ref{eq:lcs1}) to infinity without loss of generality of the power-law asymptotic expansion, owing to the exponentially fast decay of the integrand, we may first evaluate the integral by parts, second by subsequently using another change of variable $w=u^2$, and third by using the definition of the gamma function $\Gamma(z)= \int_0^\infty s^{z-1}e^{-s} ds$
\begin{align}\begin{split}
\int_0^\infty u^{n+1}\, E_1\!\left(u^2\right)du &= \frac{2}{n+2}\int_0^\infty u^{n+1} e^{-u^2} du \\[9 pt]
&=\frac{1}{n+2}\int_0^\infty w^{n/2} e^{-w} dw=\frac{\Gamma(n/2+1)}{n+2}=\frac{(n/2)!}{n+2}
\label{eq:lcs2}
\end{split}
\end{align}
Combining (\ref{eq:lcs1}) with (\ref{eq:lcs2}), and changing the index of summation $n=2m$, we find the asymptotic expansion of $T$ for large $\lambda$ in the critically stressed limit
\begin{equation}
T\sim \sum_{m=0}^\infty \frac{1}{\sqrt{\pi}}\frac{[(2m-1)/2]!}{2m+2} \frac{1}{\lambda^{2+2m}}=\frac{1}{2\lambda^2}+\frac{1}{8\lambda^4}+\frac{1}{8\lambda^6}+...\label{eq:Tcs}
\end{equation}

\

\begin{figure}
      \center\includegraphics[width=179.4pt]{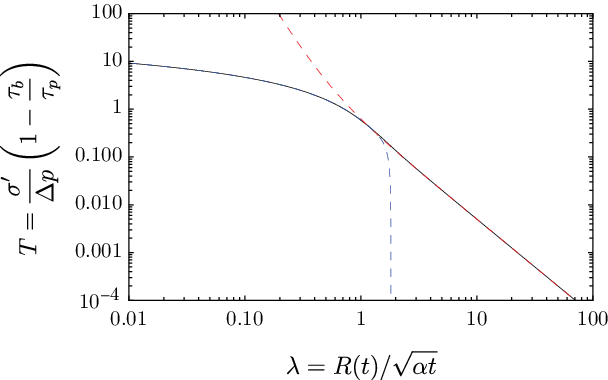}

   \caption{\raggedright (solid black) Plot of the variation of the stress-injection parameter $T$, which characterizes the pre-injection state of fault stress and strength as well as the strength of the injection, with the prefactor $\lambda$ for the self-similar expansion of the crack radius, $R(t)=\lambda \sqrt{\alpha t}$. The marginally pressurized (\ref{eq:Tmp}, blue-dashed) and critically stressed (\ref{eq:Tcs}, red-dashed) are the perturbation expansions in the small and large $\lambda$ limit, respectively. The marginally pressurized expansion is carried to $O(\lambda^8)$ and the critically stressed expansion is carried to $O(\lambda^{-4})$, which represents an optimal truncation for $\lambda$ near 1 (discussed further in section \ref{sec:opt}).}

   \label{fig:T}
\end{figure}

For small values of $\lambda$, we may evaluate (\ref{eq:fT}) using the series expansion of the exponential integral for small values of its argument [e.g., \emph{Bender and Orszag}, 1999]
\begin{equation}
E_1\!\left[(\lambda x)^2\right]= -\gamma - 2 \ln (\lambda x)-\sum_{n=1}^\infty (-1)^n \frac{(\lambda x)^{2n}}{n \cdot n!}
\end{equation}
where $\gamma=0.57721...$ is the Euler-Maraschoni constant, such that 
\begin{align}
\begin{split}
T&= - \gamma\int_0^1\frac{x}{\sqrt{1-x^2}}dx -2\int_0^1 \frac{\ln(\lambda x)}{\sqrt{1-x^2}}dx-\sum_{n=1}^\infty (-1)^n \frac{\lambda^{2n}}{n \cdot n!}\int_0^1 \frac{x^{2n +1}}{\sqrt{1-x^2}}dx\\[9 pt]
&= - \gamma + 2 -2 \ln(2\lambda) - \sum_{n=1}^\infty (-1)^n \frac{\lambda^{2n}}{n \cdot n!}\left( n! \frac{\sqrt{\pi}/2}{(n+1/2)!}\right)\\[9 pt]
&= - \gamma + 2 -2 \ln(2\lambda) - \sum_{n=1}^\infty (-1)^n \frac{\sqrt{\pi}/2}{n\cdot (n+1/2)!} \lambda^{2n}\\[9 pt]
&= - \gamma + 2 -2 \ln(2\lambda) + \frac{2}{3}\lambda^2-\frac{2}{15}\lambda^4+\frac{8}{315}\lambda^6 -...
\end{split}\label{eq:Tmp}
\end{align}
in which the first integral is readily evaluated, the second likewise but by parts, and the evaluation of the last integral is to be found as an ancillary exercise in appendix \ref{app:mp}, where we identify it as $I_{2n+1}$. In Figure \ref{fig:T} we compare the variation of $T$ with $\lambda$, satisfying (\ref{eq:fT}), against the marginally pressurized and critically stressed expansions.

\

We are interested in deriving series expansions for slip in the limits of large and small $\lambda$. In the marginally pressurized limit, $\lambda\ll1$, the rupture radius lags far behind the fluid diffusive lengthscale $\sqrt{\alpha t}$. In the critically stressed limit, where, $\lambda\gg1$, the rupture radius is much larger than the diffusive lengthscale, $R(t)\gg\sqrt{\alpha t}$, and with this scale separation, there will be an outer solution valid over distances $\sim R$ and an inner solution relevant over distances $\sim\sqrt{\alpha t}$ near $r=0$. This inner solution is in-effect, an interior boundary layer. In what follows, we present in section \ref{sec:cs} the asymptotic expansion of the inner and outer solutions for the critically stressed case, as well as their asymptotic matching and the construction of a composite solution; and we show in section \ref{sec:mp} the asymptotic expansion of slip in the marginally pressurized limit in section. Details of the solution procedure will be relegated to appendix \ref{app:in} for the inner solution, appendices \ref{app:out} and \ref{app:outa} for the outer solution, and appendix \ref{app:mp} for the marginally pressurized solution. 

\section{Critically stressed faults: inner, outer, and composite solutions}
\label{sec:cs}
For small values of the stress parameter $T$ (large $\lambda$), the critically stressed fault is highly sensitive to the injection of fluids: rupture outpaces the migration of injected fluids. We perform a singular perturbation expansion of the self-similar slip profile in terms of the small parameter $\lambda^{-1}$. In the following, we provide the behavior of slip on an interior boundary layer about the origin in section \ref{sec:is}. This solution captures behavior near the origin and will asymptotically decay away from it. Subsequently, in section \ref{sec:os}, we provide an outer solution that captures behavior at distances greater than the diffusive length away from the origin and up to the rupture front. The outer solution's approximation breaks down at the origin, where that solution is singular. We will find that the manner of the asymptotic decay of the inner solution matches the singularity of the outer solution and that the two overlap over distances intermediate to the diffusive lengthscale and rupture front position. In section \ref{sec:mae}, we construct a composite solution as the sum of the inner and outer solutions, less the common overlap. 

\subsection{Inner solution}
\label{sec:is}
To examine the variation of slip over distances of the order of the diffusive lengthscale $\sqrt{\alpha t}$ away from the origin, we use the scaled radial position $s=r/\sqrt{\alpha t}$, which has the domain $0\leq s \leq \lambda$. While the details of the solution are deferred to appendix A, the final result is that the asymptotic expansion for slip in the large $\lambda$ limit, is given by $\delta\sim\delta_{in}$ where
\begin{align}\frac{\delta_{in}(s,t)}{\sqrt{\alpha t} f \Delta p/\mu}=\delta_0(s,t)+\frac{4}{\pi^2}\sum_{j=0}^\infty\frac{1}{\lambda^{2j+1}}\frac{1}{2j+1}\left[\sum_{k=0}^j s^{2k}\frac{\Gamma(k-1/2)}{\Gamma(k+1)}\frac{\Gamma(3/2+j-k)}{2(j-k)+1}\right]\label{eq:dinf}\end{align}
for which the leading, zeroth-order term is
\begin{equation}\delta_0(s,t)=2\sqrt{\pi}e^{-s^2/2} \left[(1+s^2)I_0(s^2/2)+s^2 I_1(s^2/2)\right]-4s\label{eq:d0}\end{equation}
and $I_0$ and $I_1$ are modified Bessel functions of the first kind. The subsequent, higher-order terms are polynomial and, to fifth-order in the small parameter $1/\lambda$, the expansion is
\begin{equation}
\frac{\delta_{in}(s,t)}{\sqrt{\alpha t} f\Delta p/\mu}=\delta_0(s,t)-\frac{1}{\lambda}\frac{4}{\pi}-\frac{1}{\lambda^3}\frac{2}{3\pi}(1-s^2)-\frac{1}{\lambda^5}\frac{1}{5\pi}\left(3-s^2-s^4/2\right)+O(\lambda^{-7})
\end{equation}

\subsection{Outer solution}
\label{sec:os}
We now examine behavior over radial distances comparable to the current rupture radius $R(t)$, relatively much larger than the diffusive lengthscale $\sqrt{\alpha t}$. We rescale radial distance as $\rho=r/R(t)$, and find the asymptotic expansion of the outer solution for slip as $\lambda\rightarrow\infty$ is given by $\delta\sim\delta_{out}$ where
\begin{align}\begin{split}
\frac{\delta_{out}(\rho,t)}{R(t) f \Delta p/\mu}=\frac{4}{\pi^{2}}\sum_{n=0}^\infty&\frac{1}{\lambda^{2n+2}}\frac{1}{\rho^{2n+1}}\frac{\Gamma(n+3/2)}{2n+1}\frac{\Gamma(n+1/2)}{\Gamma(n+2)}  \\[9 pt]
&\cdot\left[\left(\text{acos}(\rho)-\rho\sqrt{1-\rho^2}\right)+2\sqrt{\pi}\frac{(1-\rho^2)^{3/2}}{\rho}\sum_{k=1}^{n}\frac{\rho^{2k}}{2k+2} \frac{\Gamma(k+2)}{\Gamma(k+1/2)}\right]
\label{eq:dout}
\end{split}\end{align}
where we have deferred discussion of the solution details to appendix B (and an alternative derivation using a multipole expansion to appendix C). This expansion, to sixth-order in the small parameter $1/\lambda$ when scaled as such, reduces to
\begin{align*}\frac{\delta_{out}(\rho,t)}{R(t) f \Delta p/\mu}=&\frac{1}{\lambda^2}\frac{2}{\pi}\left(\frac{\text{acos}(\rho)}{\rho}-\sqrt{1-\rho^2}\right)+\frac{1}{\lambda^4}\frac{1}{4 \pi}\left[\frac{\text{acos}(\rho)}{\rho^3}-\sqrt{1-\rho^2}\left(2-\frac{1}{\rho^2}\right)\right] \\[9 pt]
&+\frac{1}{\lambda^6}\frac{3}{16\pi}\left[\frac{\text{acos}(\rho)}{\rho^5}-\sqrt{1-\rho^2}\left(\frac{8}{3}-\frac{2}{3}\frac{1}{\rho^2}-\frac{1}{\rho^4}\right)\right]+O(\lambda^{-8})
\end{align*}

\

\begin{figure}
      \center\includegraphics[width=368.6pt]{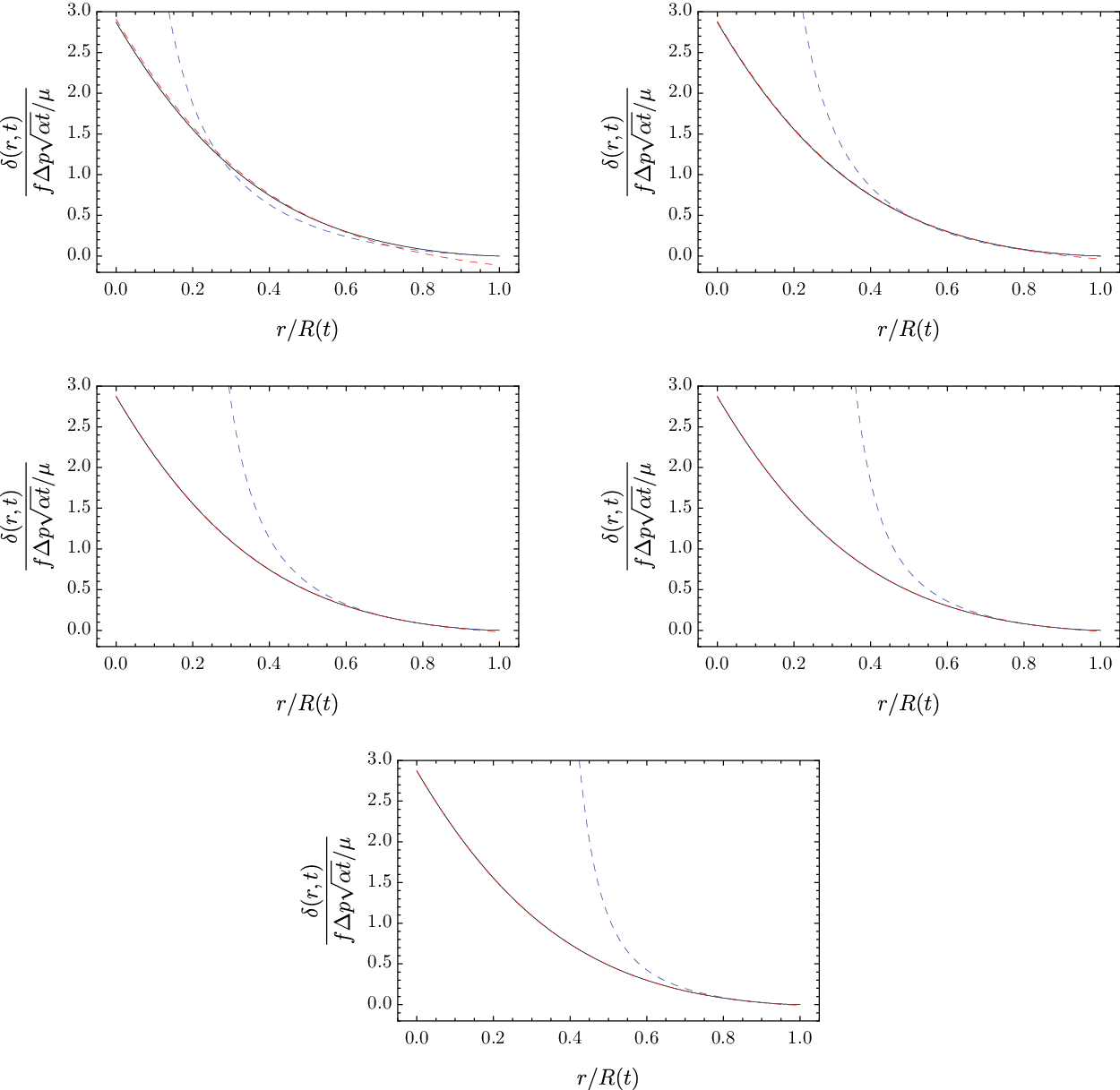}
   \caption{\raggedright Plots of inner (red-dashed) and outer (blue-dashed) solutions in comparison with full solution (solid black), for a case modestly within the critically stressed regime, $\lambda=2$. The inner and outer solutions are carried out to $O(\lambda^{-1})$ (top left) $O(\lambda^{-3})$ (top right) $O(\lambda^{-5})$ (middle left) $O(\lambda^{-7})$ (middle right) and $O(\lambda^{-9})$ (bottom). The bottommost panel shows the optimally truncated inner and outer solutions of the expansion (discussion further in section \ref{sec:opt}). }

   \label{fig:io}
\end{figure}

\begin{figure}
      \center\includegraphics[width=381.2pt]{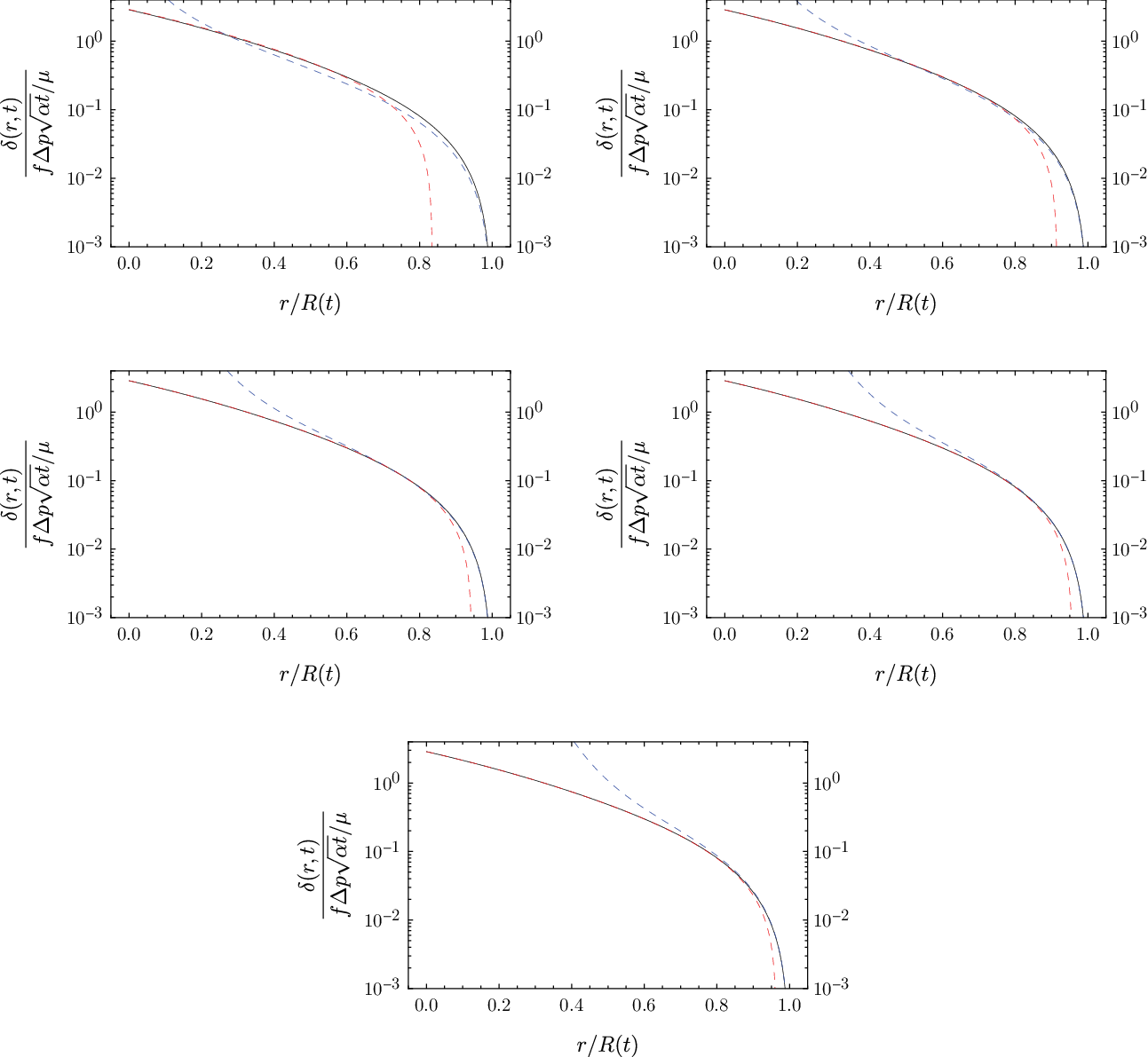}
   \caption{\raggedright Same as in Figure \ref{fig:io}, except with the ordinate axes now on a logarithmic scale to highlight separation of inner and outer solutions, as well as their diminishing overlap as order is increased to the order of optimal truncation (bottommost panel).}

   \label{fig:iolog}
\end{figure}

We may compare the inner and outer solutions with direct evaluation of the inversion for slip (\ref{eq:dt}). For a fixed value of $\lambda=2$---only slightly within the critically stressed regime---the panels of Figure \ref{fig:io} show the inner and outer solutions as red- and blue-dashed lines, respectively, overlaid on a direct solution for the self-similar slip distribution $\delta$ (solid black line). The series of panels shows the inner and outer solutions truncated to progressively greater order: i.e., the inner and outer series (\ref{eq:dinf}) and (\ref{eq:dout}), when each are scaled by $\sqrt{\alpha t} f\Delta p/\mu$, are truncated to $O(\lambda^{-(2N+1)}$) with $N$ increasing from 0 to 4 as panels progress from left to right and in descending rows. To within the range examined, the increasing order of the inner and outer asymptotic solutions show progressively better approximation, with the inner solution appearing to capture behavior over the entire crack. Plotting the ordinate axes of the panels of Figure \ref{fig:io} on a logarithmic scale (Figure \ref{fig:iolog}), we improve the visible separation of coverage of the inner and outer solutions of their respective regions, as well as the narrowing region of overlap between these solutions.

\subsection{Matched asymptotics and composite solution}
\label{sec:mae}
The inner and outer expansions can be used to construct a composite asymptotic solution, which will provide uniform convergence over the entire domain [e.g., \emph{Hinch}, 1991]. The composite solution for the critically stressed limit is given by
\begin{equation}\frac{\delta_{cs}(\rho,t)}{\sqrt{\alpha t} f\Delta p/\mu}=\frac{\delta_{in}(\lambda \rho,t)+\delta_{out}(\rho,t)-\delta_{overlap}(\rho,t)}{\sqrt{\alpha t} f\Delta p/\mu}\label{eq:comp}\end{equation}
where $\delta_{overlap}$ is the common asymptotic behavior of the inner and outer solutions in an intermediate region corresponding to the outer region ($s\rightarrow \infty$) of the inner solution, and the inner region ($\rho\rightarrow 0$) of the outer solution. This overlap is subtracted to avoid double-counting contributions in this intermediate region when summing the inner and outer solutions. 

\

The expression for the inner solution is given as before, (\ref{eq:dinf}--\ref{eq:d0}). Regarding the outer solution, we must carefully note that the  scaling for the composite solution (\ref{eq:comp}) does not match that of the prior expression of the outer solution (\ref{eq:dout}): their scalings differ by a factor of $R(t)/\sqrt{\alpha t}=\lambda$. Rescaling the outer solution in the manner of (\ref{eq:comp}),
\begin{align}\begin{split}
\frac{\delta_{out}(\rho,t)}{\sqrt{\alpha t} f \Delta p/\mu}=\frac{4}{\pi^{2}}\sum_{n=0}^\infty&\frac{\Gamma(n+3/2)}{2n+1}\frac{\Gamma(n+1/2)}{\Gamma(n+2)}\frac{1}{(\lambda\rho)^{2n+1}}  \\[9 pt]
&\cdot\left[\left(\text{acos}(\rho)-\rho\sqrt{1-\rho^2}\right)+2\sqrt{\pi}\frac{(1-\rho^2)^{3/2}}{\rho}\sum_{k=1}^{n}\frac{\rho^{2k}}{2k+2} \frac{\Gamma(k+2)}{\Gamma(k+1/2)}\right]
\label{eq:doutrs}
\end{split}\end{align}
We may determine the overlap by examining the outer ($s\rightarrow \infty$) behavior of the inner solution or the inner ($\rho\rightarrow 0$) behavior of the outer solution. Focusing on the inner solution, its far-field behavior is given by the power-law decay of the leading order term $\delta_0(s,t)$, which we determine in Appendix \ref{app:d0}, plus the polynomial terms contained in the double sum within (\ref{eq:dinf}) that are the higher-order inner solution terms. These are, respectively, the first and second sums over $n$ in the following expression for the overlap
\begin{align}\begin{split}
\frac{\delta_{overlap}(\rho,t)}{\sqrt{\alpha t} f\Delta p/\mu}=&\frac{2}{\pi}\sum_{n=0}^\infty \frac{\Gamma(n+3/2)}{2n+1}\frac{\Gamma(n+1/2)}{\Gamma(n+2)}\frac{1}{(\lambda \rho)^{2n+1}}\\[9 pt]
&+\frac{4}{\pi^2}\sum_{n=0}^\infty\frac{1}{\lambda^{2n+1}}\frac{1}{2n+1}\left[\sum_{k=0}^n (\lambda \rho)^{2k}\frac{\Gamma(k-1/2)}{\Gamma(k+1)}\frac{\Gamma(3/2+n-k)}{2(n-k)+1}\right]
\end{split}\end{align}
The composite solution is hence
\begin{align}\begin{split}
\frac{\delta_{cs}(\rho,t)}{\sqrt{\alpha t} f\Delta p/\mu}=&2 \sqrt{\pi} e^{-(\lambda \rho)^2/2} \left\{[1+(\lambda \rho)^2]I_0\left[\frac{(\lambda \rho)^2}{2}\right]+(\lambda \rho)^2 I_1\left[\frac{(\lambda \rho)^2}{2}\right]\right\} -4 \rho \lambda\\[9 pt]
&+\frac{4}{\pi^{2}}\sum_{n=0}^\infty\frac{1}{(\lambda\rho)^{2n+1}}\frac{\Gamma(n+3/2)}{2n+1}\frac{\Gamma(n+1/2)}{\Gamma(n+2)}\label{eq:cs}\\[9 pt]
&\cdot\left[\left(\text{acos}(\rho)-\rho\sqrt{1-\rho^2}\right)+2\sqrt{\pi}\frac{(1-\rho^2)^{3/2}}{\rho}\sum_{k=1}^{n}\frac{\rho^{2k}}{2k+2} \frac{\Gamma(k+2)}{\Gamma(k+1/2)}-\frac{\pi}{2}\right]
\end{split}\end{align}
The last term in brackets may be slightly simplified using the identity $\text{acos}(\rho)-\pi/2=-\text{asin}(\rho)$.

\

To illustrate the asymptotic matching, we examine the above-mentioned series truncated to three terms. In this case, the inner solution is
\begin{align*}\frac{\delta_{in}(s,t)}{\sqrt{\alpha t}f \Delta p/\mu}=\,\,&2 \sqrt{\pi} e^{-s^2/2} \left[(1+s^2)I_0(s^2/2)+s^2 I_1(s^2/2)\right] -4s \\[9 pt]
&-\frac{4}{\pi \lambda}-\frac{2}{3\pi\lambda^3}\left(1-s^2\right)-\frac{1}{5\pi\lambda^5}\left(3-s^2-\frac{s^4}{2}\right) +O(\lambda^{-7})\end{align*}
and the outer solution, rescaled in the manner of (\ref{eq:comp}), is
\begin{align*}\frac{\delta_{out}(\rho,t)}{\sqrt{\alpha t} f \Delta p/\mu}=&\frac{1}{\lambda}\frac{2}{\pi}\left(\frac{\text{acos}(\rho)}{\rho}-\sqrt{1-\rho^2}\right)+\frac{1}{\lambda^3}\frac{1}{4 \pi}\left[\frac{\text{acos}(\rho)}{\rho^3}-\sqrt{1-\rho^2}\left(2-\frac{1}{\rho^2}\right)\right] \\[9 pt]
&+\frac{1}{\lambda^5}\frac{3}{16\pi}\left[\frac{\text{acos}(\rho)}{\rho^5}-\sqrt{1-\rho^2}\left(\frac{8}{3}-\frac{2}{3}\frac{1}{\rho^2}-\frac{1}{\rho^4}\right)\right]+O(\lambda^{-8})
\end{align*}
We may now evaluate the so-called overlap of the inner and outer solutions, here both expanded to fifth-order in $\lambda^{-1}$ when scaled as such.

\

The overlap of these two solutions is determined by comparing the outer behavior of the inner solution, i.e., the asymptotic behavior of $\delta_{in}(s\rightarrow\infty,t)$ with the inner behavior of the outer solution, i.e., the asymptotic behavior of $\delta_{out}(\rho\rightarrow 0$,t). The latter can be determined to be
\begin{align*}
\frac{\delta_{out}(\rho\rightarrow0,t)}{\sqrt{\alpha t} f\Delta p/\mu}&=\frac{1}{\lambda}\left(\textcolor{red}{\frac{1}{\rho}-\frac{4}{\pi}}+\textcolor{blue}{\frac{2}{3}\frac{\rho ^2}{\pi}}+\textcolor{green}{\frac{1}{10\pi}\rho^4}+O(\rho^6)\right) \\[9 pt]
&+\frac{1}{\lambda^3}\left(	\textcolor{blue}{\frac{1}{8 \rho^3}-\frac{2}{3\pi}} +\textcolor{green}{\frac{\rho^2}{5 \pi}} + \frac{\rho^4}{28 \pi}+O(\rho^{6})		\right)\\[9 pt]
&+\frac{1}{\lambda^5}\left(\textcolor{green}{\frac{3}{32 \rho^5} -\frac{3}{5\pi}}+\frac{3 \rho^2}{14 \pi} +\frac{\rho^4}{24\pi} +O(\rho^6)		\right) +O(\lambda^{-7})
\end{align*}
where we again note here that the scaling of $\delta_{out}$ on the left-hand side is different from the scaling of slip in (\ref{eq:dout}), in which $R(t)=\lambda\sqrt{\alpha t}$ was used in place of $\sqrt{\alpha t}$. In the equation above, we scale slip with $\sqrt{\alpha t}$ so that we may correctly compare this inner behavior with the outer behavior of the inner solution, which is 
\begin{equation*}\frac{\delta_{in}(s\rightarrow\infty,t)}{\sqrt{\alpha t} f\Delta p/\mu}=\left(\textcolor{red}{\frac{1}{s}}+\textcolor{blue}{\frac{1}{8s^3}}+\textcolor{green}{\frac{3}{32s^5}}+O(s^{-7})\right) \textcolor{red}{-\frac{1}{\lambda}\left(\frac{4}{\pi}\right)} \textcolor{blue}{-\frac{1}{\lambda^3}\left(\frac{2}{3\pi}(1-s^2)\right)}\textcolor{green}{-\frac{1}{\lambda^5}\left(\frac{1}{5\pi}(3-s^2-s^4/2)\right)}\end{equation*}
Recognizing that $s=\rho\lambda$, we can identify the common, overlapping terms for the two expansions, which we color in red, blue, and green to facilitate comparisons. This overlap is clearly
\begin{align}\begin{split}\frac{\delta_{overlap}(\rho,t)}{\sqrt{\alpha t} f\Delta p/\mu}&=\textcolor{red}{\frac{1}{\lambda}\left(\frac{1}{\rho}-\frac{4}{\pi}\right)}+\textcolor{blue}{\frac{1}{\lambda^3}\left[\frac{1}{8 \rho^3}-\frac{2}{3\pi}\left(1-(\rho\lambda^2)\right)\right]}\\[9 pt]
&+\textcolor{green}{\frac{1}{\lambda^5}\left[\frac{3}{32\rho^5}-\frac{1}{5\pi}\left(3 -(\rho\lambda)^2-\frac{(\rho\lambda)^4}{2}\right)\right]}+O(\lambda^{-7})\end{split}\end{align}

The composite solution to fifth order in the small parameter $1/\lambda$ is then
\begin{align}
\begin{split}
\frac{\delta_{cs}(\rho,t)}{\sqrt{\alpha t} f\Delta p/\mu}&=2 \sqrt{\pi} e^{-(\lambda \rho)^2/2} \left\{[1+(\lambda \rho)^2]I_0\left[\frac{(\lambda \rho)^2}{2}\right]+(\lambda \rho)^2 I_1\left[\frac{(\lambda \rho)^2}{2}\right]\right\} -4\lambda \rho\\[9 pt]
&-\frac{1}{\lambda}\frac{2}{\pi}\left(\frac{\text{asin}(\rho)}{\rho}+\sqrt{1-\rho^2}\right)\\[9 pt]
&-\frac{1}{\lambda^3}\frac{1}{4 \pi}\left[\frac{\text{asin}(\rho)}{\rho^3}+\sqrt{1-\rho^2}\left(2-\frac{1}{\rho^2}\right)\right] \\[9 pt]
&-\frac{1}{\lambda^5}\frac{3}{16\pi}\left[\frac{\text{asin}(\rho)}{\rho^5}+\sqrt{1-\rho^2}\left(\frac{8}{3}-\frac{2}{3}\frac{1}{\rho^2}-\frac{1}{\rho^4}\right)\right]+O(\lambda^{-7})
\end{split}
\end{align}

\begin{figure}
      \center\includegraphics[width=376.5pt]{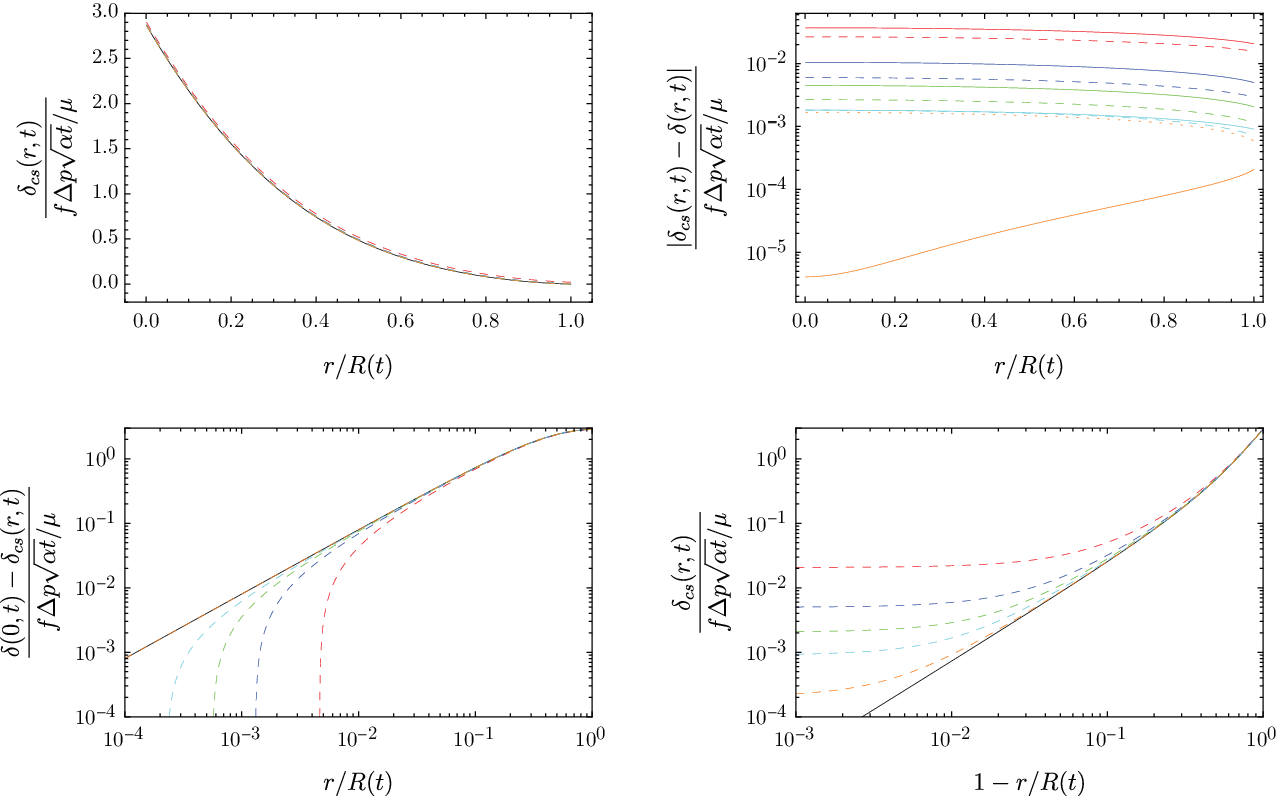}
   \caption{\raggedright ({\bf top left}) Comparison of composite solutions for slip at various orders $\delta_{cs}$ (dashed lines) against the solution for slip $\delta$ (black line) at a value of $\lambda=2$ only modestly in the critically stressed regime. Composite solutions are carried out to the following orders in the small parameter $\lambda^{-1}$: leading first (red), third (blue), fifth (green), seventh (cyan), and ninth (orange). By fifth order, the composite solution is indistinguishable from the true solution. ({\bf top right}) Absolute error of composite solutions at various orders are shown as solid line, whose colors indicate order of composite solution and correspond to those of dashed lines in top-left. Dashed and dotted lines indicate value of the next-order term in the composite solution: e.g., the red-dashed line shows that the value of the third-order ($\lambda^{-3}$) term of the composite solution is equal to more than 50\% of the error of the first-order solution (red-solid line). Note the dramatic reduction in error for the composite solution at ninth order (solid orange line), reflecting  optimal truncation of the composite solution (discussed further in section \ref{sec:opt}). ({\bf bottom left}) Deviation of the composite solution $\delta_{cs}$ at various orders (dashed lines) from the solution for slip $\delta$ near the crack center, $r=0$. Excellent agreement of optimally truncated composite solution at ninth order is again apparent. ({\bf bottom right}) Deviation of the composite solution near the crack tip, $r=R(t)$. }
   \label{fig:comp}
\end{figure}

We now examine the accuracy of the composite solution to represent slip that is only modestly in the critically stressed regime, considering an example case for which $\lambda$ is only moderately large: $\lambda=2$. In Figure \ref{fig:comp} (top left), we compare the composite solution constructed up to ninth order in the small parameter $\lambda^{-1}$ to a more accurate numerical solution for the complete slip profile. Beyond fifth order, the composite solution is indistinguishable from the numerical solution at this scale. To highlight that the composite solutions do deviate from the precise and accurate numerical solution at various orders, in Figure \ref{fig:comp} (bottom left) and \ref{fig:comp} (bottom right) we show the discrepancy from the numerical solution on a logarithmic scale at the crack center $r=0$ and crack tip $r=R(t)$, respectively. In Figure \ref{fig:comp} (top right) (solid lines) we show the distribution along the crack of the absolute error of the composite solution at various orders. We compare this error with the value of the next-order term in the composite solution (dashed lines). We see that by the seventh-order solution (solid cyan), much of the error is contained in the next-order term (underlying dashed cyan). At ninth order, we see a dramatic drop in error, with the next-order term (dotted orange) having a much larger value. This is an indication that, for $\lambda=2$, the composite solution is most accurate at ninth order. Carrying out the composite solution to higher order will only increase the error. We postpone further discussion of optimal truncation to section \ref{sec:opt}. We see that even for mildly critically stressed conditions, the composite solution is capable of achieving absolute errors of the order of $10^{-5}$, small in comparison with the $O(1)$ value of slip along the crack.

\section{Fault slip under marginally pressurized conditions}
\label{sec:mp}
In the limit of large $T$ (small $\lambda$), the self-similar fluid pressure profile stretches faster than crack growth: rupture lags the injected fluid. In this case, we may perform a regular perturbation expansion for the slip profile in powers of the small parameter $\lambda$. The details of the solution may be found in appendix \ref{app:mp}, but the perturbation to arbitrary order is given by

\begin{equation}
\frac{\delta_{mp}(\rho,t)}{R(t)f \Delta p/\mu}=\frac{8}{\pi}\left(\sqrt{1-\rho^2}-\rho\text{ acos}(\rho)\right)+2\left(1-\rho^2\right)^{3/2}\sum_{n=1}^\infty \frac{\lambda^{2n}}{2n+1}\frac{(-1)^n}{\Gamma(n+3/2)}\sum_{k=1}^{n}\frac{\Gamma(k)}{\Gamma(k+1/2)}\rho^{2k-2}
\label{eq:mpf}
\end{equation}
To $O(\lambda^4)$, the asymptotic solution is
\begin{equation*}
\frac{\delta_{mp}(\rho,t)}{R(t)f \Delta p/\mu}=\frac{8}{\pi}\left(\sqrt{1-\rho^2}-\rho\text{ acos}(\rho)\right)-\lambda^2\frac{16}{9\pi}(1-\rho^2)^{3/2}+\lambda^4\frac{32}{225\pi}(1-\rho^2)^{3/2}(3+2\rho^2)+O(\lambda^6)
\end{equation*}

\begin{figure}
	\center\includegraphics[width=372.4pt]{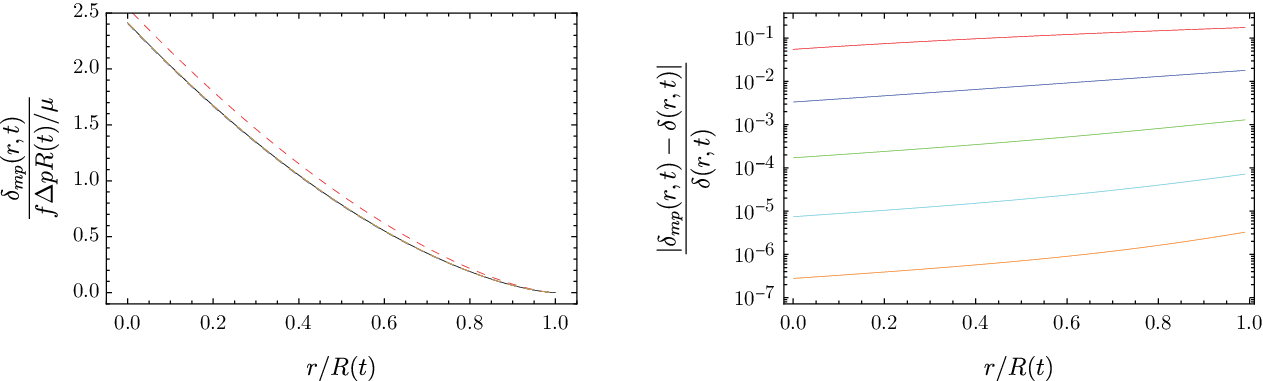}
   \caption{\raggedright({\bf left}) Comparison of the marginally pressurized series solution truncated at various orders (dashed-lines) with the full solution (black solid line), both for $\lambda=1/2$. The series are carried up to: $O(\lambda^0)$ (red), $O(\lambda^2)$ (blue), $O(\lambda^4)$ (green), $O(\lambda^6)$ (cyan), and $O(\lambda^8)$ (orange). Beyond the leading order, the marginally pressurized solutions are indistinguishable at this scale. ({\bf right}) Plot of the relative error of the truncated series solutions, with the same color scale as the right panel. For the regular perturbation series exhibits uniform convergence in terms of relative error.}
   \label{fig:mp}
\end{figure}

In Figure \ref{fig:mp} we examine a mildly marginally pressurized solution ($\lambda=1/2$). The top-left panel compares the full solution (solid black curve) with the marginally pressurized solution truncated to various orders: $O(\lambda^0)$, $O(\lambda^2)$, $O(\lambda^4)$, and $O(\lambda^6)$. The top-right panel (panel 2) shows relative error at each order. In contrast with the critically stressed case, in which the absolute error showed uniform convergence, here it is the relative error that is uniform, such that the divergence of the type seen in, for example, the bottom right panel of Figure \ref{fig:comp} does not occur for the truncated marginally pressurized solutions. The regularly decreasing error with increase in order in the series expansions exhibits the standard convergence expected for a regular perturbation expansion. We discuss this convergence further in section \ref{sec:reg}.

\section{Aseismic moment release; peak slip at injection point}
\label{sec:Md0}
We now calculate two quantities, one characterizing far-field behavior and the other local, as they each depend on $\lambda$ in the marginally pressurized and critically stressed limits. The first quantity, is the so-called aseismic moment, which is the leading-order moment in a multipole expansion of the far-field behavior (e.g, displacement) due to fault slip. However, we also calculate a near-field quantity: the peak slip, which occurs at the crack center.

\

\begin{figure}
	\center\includegraphics[width=370.8pt]{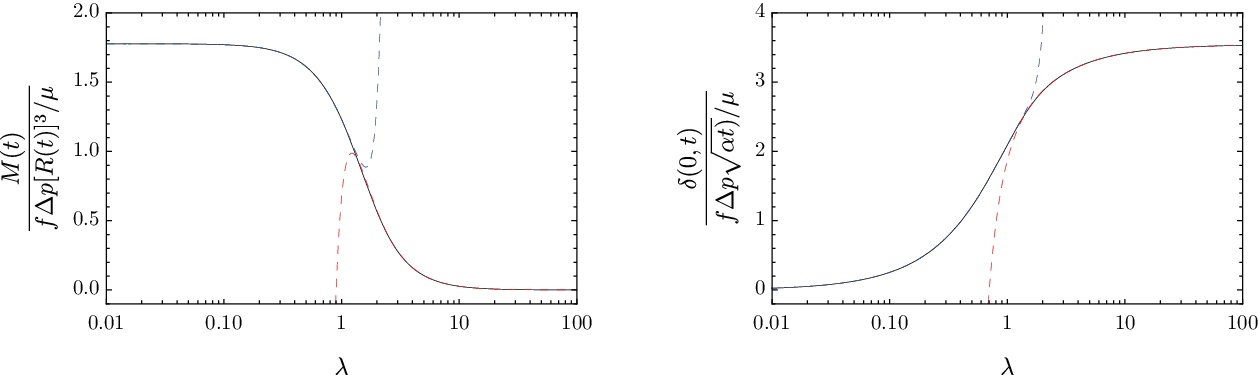}
   \caption{\raggedright Comparison of aseismic moment ({\bf left}) and peak slip ({\bf right}) as a function of $\lambda$ (solid black) compared with truncated marginally pressurized (blue-dashed) and critically stressed (red-dashed) expansions. ({\bf left}) The marginally pressurized expansion is $O(\lambda^{8})$ and the critically stressed expansion is $O(\lambda^{-6})$. The former reflects the first four terms of (\ref{eq:Mmp}); the latter reflects the first three terms of (\ref{eq:Mcs}) and a $\lambda^{-1}$ scaling factor due to a change in normalization. ({\bf right}) The marginally pressurized expansion $\delta_{mp}$ is $O(\lambda^9)$ and the critically stressed expansion $\delta_{cs}$ is $O(\lambda^{-5})$. The former reflects the first five terms of (\ref{eq:psmp}) and a $\lambda$ scaling factor due to a change in normalization. The latter reflects the first three terms of (\ref{eq:pscs}). In both panels, truncation of the critically stressed expansions at three terms is optimal for values of $\lambda$ close to 1. Optimal truncation is discussed in section \ref{sec:opt}.}
   \label{fig:Md0}
\end{figure}

The aseismic moment  is given by [e.g., \emph{Rice, 1980}]
\begin{equation*}M(t)=\mu \int_0^{R(t)} \delta(r,t) 2\pi rdr \end{equation*}
For a marginally pressurized fault, we substitute $\delta=\delta_{mp}$ in this expression for $M$ and find that 
\begin{align}\frac{M(t)}{R(t)^3 f\Delta p}&=\frac{16}{9}+\frac{3\pi^{3/2}}{2}\sum_{n=1}^\infty \frac{\lambda^{2n}}{2n+1} \frac{(-1)^n}{\Gamma(n+3/2)}\sum_{k=1}^n \frac{\Gamma(k)^2}{\Gamma(k+1/2)\Gamma(k+5/2)}\label{eq:Mmp}\\[9 pt]
&=\frac{16}{9}\left(1- \frac{2}{5}\lambda^2+\frac{4}{35}\lambda^4 - \frac{9512}{385875}\lambda^6+ O(\lambda^{8})\right)\notag
\end{align} 
For a critically stressed fault, we substitue $\delta=\delta_{cs}$ and find that 
\begin{align}\frac{M(t)}{R(t)^2 \sqrt{\alpha t} f\Delta p }&\sim e^{-\lambda^2/2}\frac{2\pi^{3/2}}{3}\left[(3+2\lambda^2)I_0(\lambda^2/2)+(1+2\lambda^2)I_1(\lambda^2/2)\right]-\frac{8\pi}{3}\lambda\notag\\[9 pt]
&+\frac{8}{\pi}\sum_{n=0}^\infty\frac{1}{\lambda^{2n+1}}\frac{\Gamma(n+3/2)}{2n+1}\frac{\Gamma(n+1/2)}{\Gamma(n+2)} \notag\\[9 pt]
&\cdot \int_0^1 \frac{1}{\rho^{2n+1}}\left[-\left(\text{asin}(\rho)+\rho\sqrt{1-\rho^2}\right)+2\sqrt{\pi}\frac{(1-\rho^2)^{3/2}}{\rho}\sum_{k=1}^{n}\frac{\rho^{2k}}{2k+2} \frac{\Gamma(k+2)}{\Gamma(k+1/2)}\right]
d\rho \notag\\[9 pt]
&\sim\frac{8}{3}\lambda^{-1}-\frac{4}{3}\lambda^{-3}-\frac{2}{3}\lambda^{-5}-\lambda^{-7}-\frac{5}{2}\lambda^{-9} -\frac{35}{4}\lambda^{-11} +O(\lambda^{-13})\label{eq:Mcs}
\end{align} 
In Figure \ref{fig:Md0} (left), we plot the variation of moment release with $\lambda$ and compare with the above-mentioned perturbation expansions. In Figure \ref{fig:Md0} (left) we normalize the moment release in the manner of (\ref{eq:Mmp}); multiplying (\ref{eq:Mcs}) by a factor of $\lambda^{-1}$ normalizes the moment release in the critically stressed limit in the manner of (\ref{eq:Mmp}).  If we instead multiply (\ref{eq:Mcs}) by a factor of $\lambda$, we find that to leading order, $M(t)\sim (8/3) \lambda (\alpha t)^{3/2} f \Delta p$.

\

We may likewise provide asymptotic expressions for the maximum slip of the rupture, $\delta(0,t)$. For critically stressed conditions, we may obtain the peak slip directly from the inner solution by evaluating expression (\ref{eq:dinf}) for $\delta_{in}(0,t)$, 
\begin{align} \frac{\delta(0,t)}{\sqrt{\alpha t} f \Delta p/\mu}&\sim 2\sqrt{\pi}\left(1-\frac{4}{\pi^2}\sum_{j=0}^\infty \frac{1}{\lambda^{2j+1}}\frac{\Gamma(3/2+j)}{(1+2j)^2} \right)\label{eq:pscs}\\[9 pt]
&\sim 2\sqrt{\pi} - \frac{4}{\pi}\frac{1}{\lambda}-\frac{2}{3\pi}\frac{1}{\lambda^3}-\frac{3}{5\pi}\frac{1}{\lambda^5}-\frac{15}{14\pi}\frac{1}{\lambda^7}-\frac{35}{12\pi}\frac{1}{\lambda^{9}}-\frac{945}{88\pi}\frac{1}{\lambda^{11}}+O(\lambda^{-13})\notag\end{align}
For marginally pressurized conditions, we  obtain an expression for the peak slip directly from (\ref{eq:mpf})
\begin{align} \frac{\delta(0,t)}{R(t) f \Delta p/\mu}&=\frac{4}{\sqrt{\pi}}\sum_{n=0}^\infty \frac{\lambda^{2n}}{2n+1}\frac{(-1)^n}{\Gamma(n+3/2)} \label{eq:psmp}\\[9 pt]
&= \frac{8}{\pi}\left(1-\frac{2}{9}\lambda^2+\frac{4}{75}\lambda^4-\frac{8}{735}\lambda^6 + O(\lambda^{8})\right)\notag
\end{align}
In Figure \ref{fig:Md0} (right), we plot the variation of the peak slip with $\lambda$ and compare with the above-mentioned perturbation expansions.

\section{Perturbation series error estimation}
\label{sec:ee}
In the preceding section, carrying out the asymptotic expansion for moment release and peak slip to higher orders makes apparent the different nature of the perturbation expansions under critically stressed and marginally pressurized conditions. For the latter, the expansion is regular and the coefficients in the series diminish in such a manner that the series is convergent, as may be shown by a ratio test. For the former, the perturbation is singular and the series is divergent.

\subsection{Optimal truncation of divergent series under critically stressed conditions}
\label{sec:opt}

In spite of the growing coefficients, we may find an optimal truncation of the divergent series. Specifically, we make use of the uniform convergence, with respect to radial distance $r$, of the composite solution to freely choose one point along crack in order to examine the diminishing benefit of keeping additional terms in the series. For convenience, we examine the asymptotic behavior of slip at the crack center. We may abstract the expression (\ref{eq:pscs}) for the slip there as
\begin{equation}
\frac{\delta(0,t)}{\sqrt{\alpha t} f \Delta p/\mu}\sim 2\sqrt{\pi} + \sum_{j=0}^\infty  a_j\lambda^{-(2j+1)}
\label{eq:trunc}
\end{equation}
and note that the series coefficients diverge for large $j$ in the manner
\begin{equation*}\frac{a_{j+1}}{a_j}\sim j-\frac{1}{2} +O(j^{-1})\end{equation*}
Comparing sequential terms we are interested in keeping terms provided that, for a given $\lambda$, the ratio of sequential terms is diminishing, or
\begin{equation*}\frac{a_{j+1}\lambda^{-(2j+3)}}{a_j\lambda^{-(2j+1)}}<1\end{equation*}
Using the leading asymptotic behavior of the coefficient ratio and rearranging, this is equivalent to the requirement that
\begin{equation}j<\lambda^2+1/2\label{eq:opt}\end{equation}
In other words, to provide the optimal approximation for the peak slip using the series expansion in (\ref{eq:pscs}) with summation index $j$ (or the optimal approximation of the entire slip profile using the composite solution (\ref{eq:cs}) with summation index $n$) the series should be carried out to the term whose summation index is the largest integer less than $\lambda^2$+1/2.

\

\begin{figure}
\center\includegraphics[width=374.6pt]{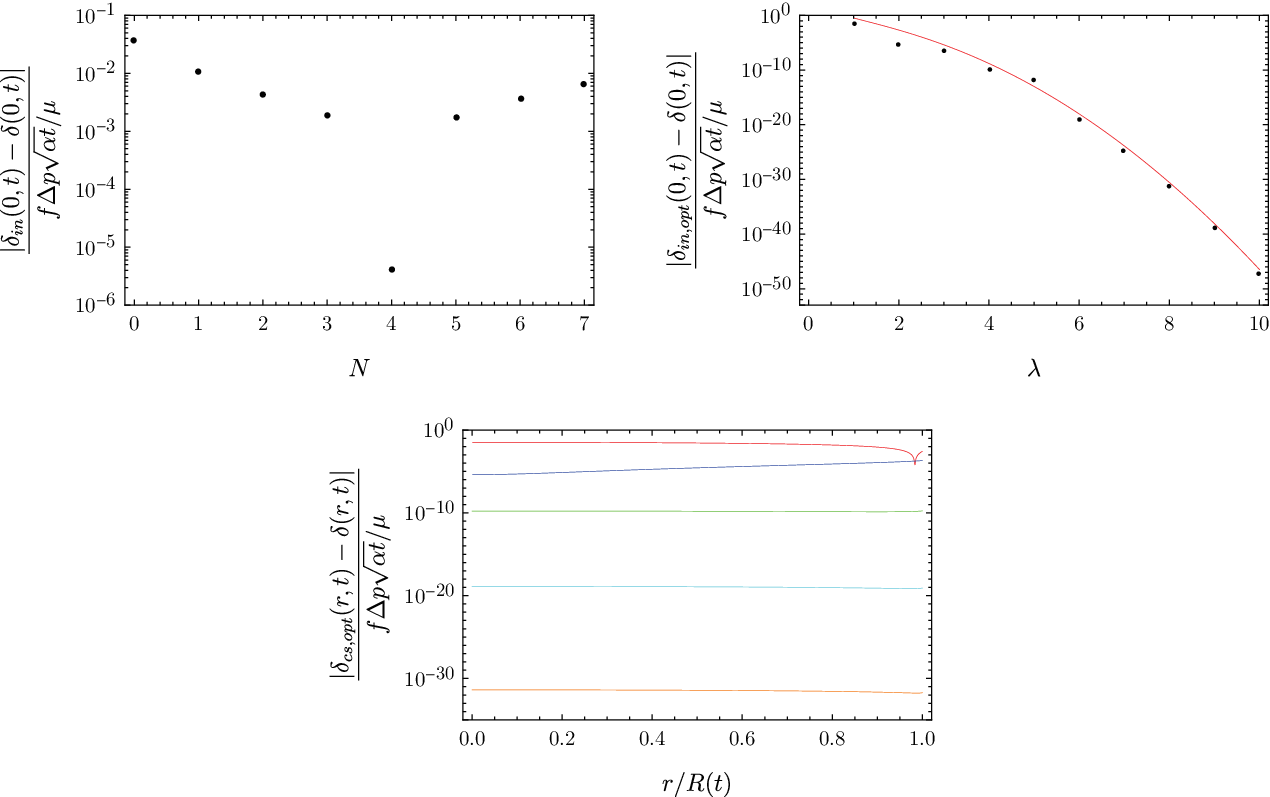}
   \caption{\raggedright {\bf (top left)} Plot of absolute error at the crack center of the asymptotic series solution (\ref{eq:pscs}) truncated at $j=N$, for $\lambda=2$. The optimal truncation is expected to be $N=\lambda^2$, and we observe a dramatic drop in error for $N=4$. {\bf (top right)} Absolute error of the composite solution at the crack center optimally truncated at $N=\lambda^2$, plotted for fixed values of $\lambda$ (black dots). The expected dependence of the error on $\lambda$ (\ref{eq:optd0}) is shown as a solid line. {\bf (bottom)} Spatial distribution of absolute error showing uniform convergence for five composite solutions (\ref{eq:cs}) optimally truncated at $N=\lambda^2$: $\lambda=1$ (red), $\lambda=2$ (blue), $\lambda=4$ (green), $\lambda=6$ (cyan), and $\lambda=8$ (orange).}
   \label{fig:opt}
\end{figure}

We may approximate the error of this optimal truncation in the critically stressed regime. We may approximate this error as the remainder left if we truncate the series (\ref{eq:trunc}) at a large value of $j$. Asymptotically, we can approximate this remainder as $a_j$, with the choice of optimal truncation $j\sim\lambda^2$. Recalling that $a_j$ stands in for terms of the series (\ref{eq:pscs}), this error estimate $a_{\lambda^2}$ evaluates as
\begin{equation*}2\sqrt{\pi}\frac{4}{\pi^2}\frac{1}{\lambda^{2 \lambda^2+1}}\frac{\Gamma(3/2+\lambda^2)}{(1+2 \lambda^2)^2}\end{equation*}
 which we may simplify using the Stirling-like formula $\Gamma(3/2+x)\sim \sqrt{2\pi} x (x/e)^x $ for large $x$ to
\begin{equation}\frac{2^{3/2}}{\pi}\frac{e^{-\lambda^2}}{\lambda^3}\label{eq:optd0}\end{equation}
which shows the optimal truncation to be exponentially accurate for large $\lambda$. 

\

We confirm exponential accuracy of the asymptotic approximations in the critically stressed limit by comparisons with analytical or sufficiently accurate numerical solutions. Examining slip at the crack center, $r=0$, an analytical expression for slip follows from substituting $s=0$ into (\ref{eq:da}),
\setlength\arraycolsep{1pt}
\begin{equation}
\frac{\delta(0,t)}{\sqrt{\alpha t} f \Delta p/\mu}=\frac{8}{\pi} \int_0^\lambda \int_0^1 \frac{x e^{-(x\xi)^2}}{\sqrt{1-x^2}}\,dx\,d\xi=\frac{8}{\pi} 
\,{}_2 F_2\!\left(\begin{matrix}&1/2& &1&\\&3/2&&3/2&\end{matrix};-\lambda^2\right)\lambda\label{eq:psan}
\end{equation}
where ${}_pF_{q}$ is the generalized hypergeometric function. We use the difference between the composite solution and this analytical solution to determine the composite solution's absolute error. We first evaluate the absolute error of the composite solution at the crack center for the particular case $\lambda=2$, truncating the series (\ref{eq:dinf}) for $\delta_{in}$ at $j=N$ where $N$ is varied. The results are shown in Figure \ref{fig:opt} (top left). The error decreases monotonically until  $N$ approaches $\lambda^2=4$, at which point a substantial decrease in error occurs and including further terms in the asymptotic expansion results in a worsening approximation. This optimal series being that truncated with $N=\lambda^2$ is consistent with the anticipated condition (\ref{eq:opt}). We now look to confirm the exponential accuracy of the asymptotic approximations as $\lambda$ increases. To do so, we evaluate the absolute error of the optimally truncated composite solution at the crack center, truncating the series expression (\ref{eq:dinf}) for $\delta_{in}$ when the summation index $j=N$, where $N=\lambda^2$. We dub this optimally truncated expansion $\delta_{in,opt}(0,t)$. We take the difference between this  and the above expression for $\delta(0,t)$ to find the absolute error at fixed values of $\lambda$, with results shown in Figure \ref{fig:opt} (top right; black dots). Comparing this to the approximated error (\ref{eq:optd0}) (Figure \ref{fig:opt} red line), we find excellent agreement. Now to examine the error over the entire crack at fixed values of $\lambda$, we first truncate the series of the composite solution (\ref{eq:cs}), expressing it up to $n=\lambda^2$ and dubbing this optimal solution $\delta_{cs,opt}$. Subtracting this from the slip $\delta$, we obtain the absolute error across the crack. Repeatedly calculating this error at fixed values of $\lambda$, we find that the error first calculated at the crack center (Figure \ref{fig:opt}, top right) is essentially uniform across the crack (Figure \ref{fig:opt}, bottom), demonstrating uniform convergence and exponential accuracy of the composite solution.

\begin{figure}
	\center\includegraphics[width=373.7pt]{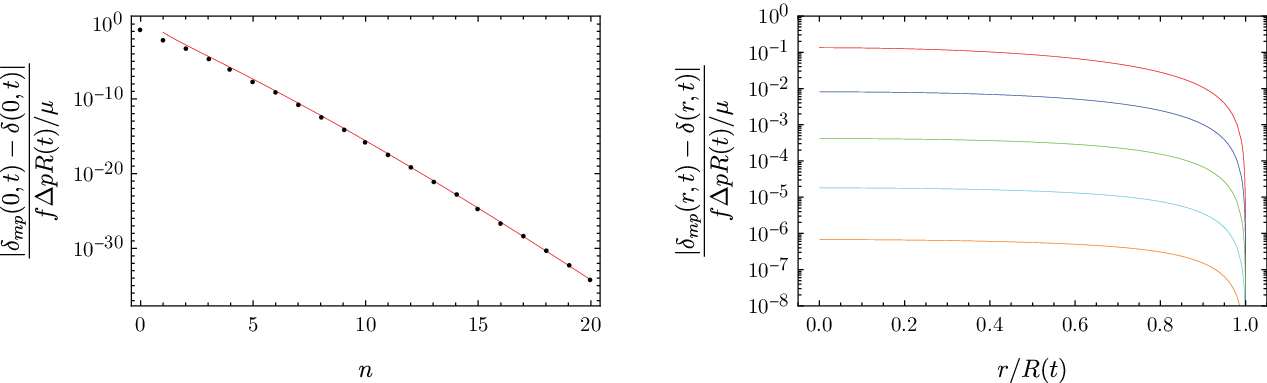}
   \caption{\raggedright({\bf left}) (black dots) Absolute error of the slip at the crack center for the marginally pressurized series solution at $O(\lambda^{2n})$ for the particular case $\lambda=1/2$. The expected error (\ref{eq:errd0}) is shown as a solid line. ({\bf right}) Distribution of the absolute error along the expanding crack for the case $\lambda=1/2$ for truncated at $O(\lambda^0)$ (red),  $O(\lambda^2)$ (blue),  $O(\lambda^4)$ (green),  $O(\lambda^6)$ (cyan),  $O(\lambda^8)$ (orange).}
   \label{fig:mperr}
\end{figure}

\subsection{Convergence of regular perturbation under marginally pressurized conditions}
\label{sec:reg}

To estimate the absolute error of truncating the series expansion of the solution under marginally pressurized conditions, we begin with the expression for the peak slip (\ref{eq:psmp}),  abstracted to 
\begin{equation}
\frac{\delta(0,t)}{R(t) f \Delta p/\mu}= \sum_{j=0}^\infty  b_j\lambda^{2j}
\end{equation}
Given that the series is a regular, convergent expansion, we may estimate the error of truncating this series at $j=n$, by the value of the $j=(n+1)$-th term. We may approximate this term for large $n$, noting that in this limit
\begin{equation*}b_{n+1}\sim\frac{\sqrt{2}}{\pi}\frac{e^n}{n^{n+3}}\end{equation*}
such that the error estimate is
\begin{equation}
\frac{\sqrt{2}}{\pi}\frac{e^n}{n^{n+3}}\lambda^{2n+2}
\label{eq:errd0}
\end{equation}
In Figure \ref{fig:mperr} (left panel), we calculate the absolute error incurred by truncating the series (\ref{eq:psmp}) at $j=n$ by subtracting this value from the analytical expression (\ref{eq:psan}), and comparing this error with the above-mentioned estimate. We confirm that, even for conditions only modestly in the marginally pressurized regime, the value of the error decreases rapidly with an increasing number of terms. Indeed, keeping only two terms in the expansion is sufficient for an absolute error of $10^{-3}$. In Figure \ref{fig:mperr} (right panel), we plot the absolute error over the extent of the crack for various truncations of the series of the marginally pressurized series solution, $\delta_{mp}$, where we see the absolute error is greatest at the crack center. From Figure \ref{fig:mp} (top right), in which we previously plotted the relative error for the same mildly pressurized case ($\lambda=1/2$), we found that the relative error is greatest at the crack tip, has a maximum of approximately 18\% for the zero-th order solution, and dramatically decreases to 1\% or 0.1\% when keeping one or two additional terms (i.e., up to second- or fourth-order in $\lambda$), respectively. 
\section{Discussion}\label{sec:dis}
\subsection{Choice of perturbation parameter}In the preceding, the perturbation parameter has been $\lambda$, the ratio of the crack radius $R$ to the diffusion lengthscale $\sqrt{\alpha t}$. This may be a quantity one may approximately infer from field measurements, say by seismological observation of seismicity migration suspected to be caused by fluid-induced aseismic slip, in which case one could in turn infer the corresponding value of initial state of fault-resolved tractions and strength of injection, reflected in the parameter $T$. However, there may be instances when $\lambda$ is not a quantity one would necessarily expect to know a priori. Rather, if one is given field measurements or inferences of stress state and fluid pressure, one would instead be in a position to calculate a value for $T$ and then look for the implied value of $\lambda$. How might one do so? A straightforward route would be to estimate $\lambda$, given $T$, by tabulating numerical evaluations of the asymptotic expansions (\ref{eq:Tcs}) or (\ref{eq:Tmp}), which can be truncated to sufficiently match the true solution over the entire parameter range (e.g., as in Figure \ref{fig:T}). This would circumvent a direct numerical solution of (\ref{eq:fT0}) or a cumbersome inversion of the hypergeometric series of \emph{S\'aez et al.} [2022]. However, when looking for expressions of slip and moment release, $\lambda$ is the natural perturbation parameter: given the initial problem formulation, one may directly proceed to find the asymptotic behavior for $T$ and slip $\delta$ as series expansions in powers of $\lambda$, as demonstrated.

\

One may nonetheless ask whether asymptotic expansions could be conducted in terms of $T$ rather than $\lambda$. Of the two quantities, $\lambda$ and $\delta$, it is only necessary to look for such expansions of the former: once a value of $\lambda$ is approximated for a given $T$, one may proceed with the provided expansions for slip and moment release in terms of $\lambda$ to the desired accuracy. With this in mind, we provide asymptotic expansions of $T$ in terms of $\lambda$ in the following.

\

In the critically stressed limit, we may look for an asymptotic expansion of  $\lambda$ as the power series $\lambda \sim \sum_{n=-1}^\infty A_{2n+1} u^{2n+1}$, where $u=\sqrt{2T}$, which we may in turn substitute for $\lambda$ in (\ref{eq:fT0}),
\begin{equation*}
T=\int_0^1 \frac{x\, E_1\!\left[\left(A_{-1}u^{-1}+A_1 u +A_3 u^3+A_5 u^5 ...\right)^2 x^2\right]}{\sqrt{1-x^2}}dx
\end{equation*}
After asymptotically approximating the integral in the limit $u\rightarrow0$ (i.e., $T\rightarrow 0$), we may successively solve for the coefficients $A_n$ by matching each side of the above at increasing order of $u$. To the leading four orders, this reduces to
\begin{align*}
O(u^2)\!:  \quad 1&=\frac{1}{A_{-1}} \\[9 pt]
O(u^4)\!: \quad 0&=1-8A_{-1}A_1\\[9 pt]
O(u^6)\!: \quad 0&=1-4 \left(A_{-1}A_1-3A_{-1}A_{1}^2+2A_{-1}^3 A_3\right)\\[9pt]
O(u^8)\!: \quad 0&=15-16\left(3A_{-1}A_1-5A_{-1}^2A_1^2+8 A_{-1}^3A_1^3\right.\\[9pt]
&\quad\left.+\,2A_{-1}^3A^{}_3-12A_{-1}^4A^{}_1A^{}_3+4A_{-1}^5A^{}_5\right)
\end{align*}
from which we find that $A_{-1}=1$, $A_1=1/8$, $A_3=11/128$, and $A_5=149/1024$ providing for the expansion of $\lambda$ in the critically stressed limit, $T\rightarrow 0$,
\begin{equation}\lambda\sim\frac{1}{\sqrt{2T}}+\frac{1}{8}\sqrt{2T} +\frac{11}{128}\sqrt{2T}^3+\frac{149}{1024}\sqrt{2T}^5+O(T^{7/2})\label{eq:Lcs}\end{equation}
a divergent series, like the expansion of $T$ in terms of $\lambda$ in the limit of large $\lambda$.

\

In the marginally pressurized limit, we may look for an expansion of $\lambda$ as the series $\lambda=\sum_{n=0}^\infty B_{2n+1}w^{2n+1}$ where $w=(1/2)\exp[(2-\gamma-T)/2]$, to be substituted for $\lambda$ in (\ref{eq:fT0}),
\begin{equation*}
T=\int_0^1 \frac{x\, E_1\!\left[\left(B_1 w+B_3w^3+B_5 w^5 +...\right)^2 x^2\right]}{\sqrt{1-x^2}}dx
\end{equation*}
This time, we may approximate the integral in the limit $w\rightarrow 0$ (i.e., $T\rightarrow \infty$) and successively solve for the coefficients $B_n$ by matching each side of the above at increasing order of $w$. To the leading four orders, this reduces to
\begin{align*}
O(\log w)\!: & \quad 0=\log(B_1) \\[9 pt]
O(w^2)\!:& \quad 0=B_1^3-3B_3 \\[9 pt]
O(w^4)\!:& \quad 0=2B_1^6-5\left(4B_1^3B_3+3 B_3^2-6B_1B_5\right)\\[9 pt]
O(w^6)\!:& \quad 0=4B_1^9-84B_1^6B_3+105 \left(B_1^3B_3^2-B^3_3+2B_1^4B_5+3B_1B_3B_5-3B^2_1B_7\right)
\end{align*}
from which we find that $B_1=1$, $B_3=1/3$, $B_5=19/90$, and $B_7=181/1134$, providing for the regular expansion of $\lambda$ in the marginally pressurized limit, $T\rightarrow\infty$,
\begin{equation}
\lambda= \frac{1}{2}e^{\frac{1}{2}(2-\gamma-T)}+\frac{1}{24}e^{\frac{3}{2}(2-\gamma-T)}+\frac{19}{2880}e^{\frac{5}{2}(2-\gamma-T)}+\frac{181}{145152}e^{\frac{7}{2}(2-\gamma-T)}+O(e^{-\frac{9}{2}T})
\label{eq:Lmp}\end{equation}

\begin{figure}
	\center\includegraphics[width=179.0pt]{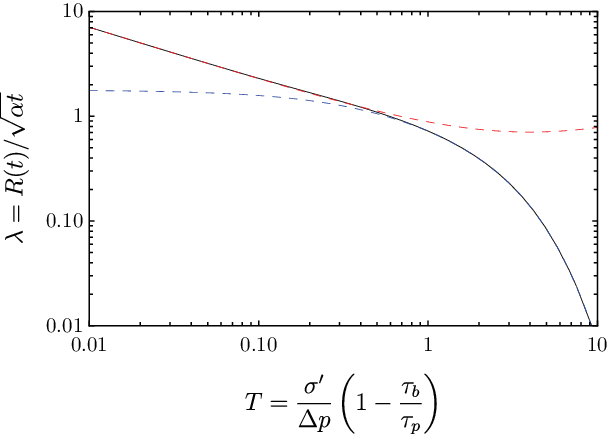}
   \caption{\raggedright  (solid black) Plot of the variation of the prefactor $\lambda$ for the self-similar expansion of the crack radius, $R(t)=\lambda \sqrt{\alpha t}$, as it depends on the stress-injection parameter $T$, which characterizes the pre-injection state of fault stress and strength as well as the strength of the injection. This plot is the inverse of Figure \ref{fig:T} and showcases the asymptotic expansions of $\lambda$ in terms of $T$: in comparison are the critically stressed (\ref{eq:Lcs}, red-dashed) and the marginally pressurized (\ref{eq:Lmp}, blue-dashed) perturbation expansions in the small and large $T$ limit, respectively. The marginally pressurized expansion shown here is carried to $O(e^{-\frac{7}{2}T})$ and the critically stressed expansion is carried to $O(T^{\frac{1}{2}})$, which represents an optimal truncation for $T$ near 1.}
   \label{fig:L}
\end{figure}

\subsection{Why $\nu=0?$}  The particular case studied here, in which Poisson's ratio $\nu=0$, may seem a peculiar, unphysical basis; however, this solution can be seen as the more readily solvable leading-order asymptotic solution for the more general problem of $\nu\neq 0$. Rupture propagation when $\nu\neq0$ is no longer axisymmetric and a free-boundary problem needs to be resolved, determining the boundary shape of the crack. This point was discussed in the preceding work of \emph{S\'aez et al.} [2023], who used a boundary-element method to explore solutions for $\nu$ in the range $0\leq\nu\leq 0.45$. They observed in their numerical solutions that the growth of rupture appears to remain self-similar when $\nu\neq0$, which itself may be unsurprising considering that the only length scale in the problem is $\sqrt{\alpha t}$. Moreover, they observed that the self-selected shape of the expanding rupture was well approximated by an ellipse over much of the parameter space, denoting the expanding length of the major and minor axes as $2a(t)$ and $2b(t)$, respectively. Remarkably, they observed that, for a given value of $T$, the area of this self-similarly expanding ellipse agreed precisely with the area of the self-similar expanding circular crack that occurs for $\nu=0$: i.e., to within an error of 1\% or less, they observed that $\pi R(t)^2=\pi a(t) b(t)$ for a given $T$ and $\sqrt{\alpha t}$, independent of the value of $\nu$. This permitted the generalization of the pre-factor $\lambda$ from the $\nu=0$ definition ($\lambda = R(t)/\sqrt{\alpha t}$) to the cases where $\nu\neq 0$, ($\lambda= \sqrt{a(t)b(t)}/\sqrt{\alpha t}$): the solution of $\lambda(T)$ from the axisymmetric $\nu=0$ was precisely that $\lambda(T)$ for the non-axisymmetric case. In addition, \emph{S\'aez et al.} [2023] observed in their numerical solutions that the ratio of major to minor axes behaved asymptotically as $a/b\sim(1-\nu)^{-1}$ and $(3-\nu)/(3-2\nu)$ in the critically stressed and marginally pressurized limits, respectively, over the explored range of $0.05\leq\nu<0.45$.

\

Thus \emph{S\'aez et al.} [2023] demonstrated that the circular crack solution adequately describes the rupture area and extent, to within a $\nu$-dependent shape correction in these limits. That and subsequent work has yet to examine, however, how the distribution of slip within the elliptic rupture ($\nu\neq0$) differs from that of a circular one ($\nu=0$) at the same $T$, e.g., whether a mapping of the axisymmetric slip from a circular to elliptical domain would suffice. Subsequent work has however, indicated that the effect of $\nu\neq 0$ on the slip is sufficiently small such that the $\nu=0$ solution can be considered an adequate leading-order solution in a perturbation expansion of the solution about $\nu=0$. Specifically, \emph{S\'aez et al.} [2025] numerically solved for the total moment released following the arrest of the self-similarly expanding aseismic rupture, due to a sudden arrest of an otherwise constant injection rate. Moment release is a useful global measure of the slip as, to within a factor of the shear modulus, it is defined as the integral of slip over the rupture area (see section 5). They found that the moment released for a non-circular rupture ($\nu=0.25$) was consistently 14\% less than circular rupture ($\nu=0$) at the same value of $T$, explored numerically over two orders of magnitude of $T$, again implying a $\nu$-dependent correction to the $\nu=0$ solution.

\

Thus, the axisymmetric, $\nu=0$ solutions represent a fairly accurate representation of the more general $\nu\neq0$ problem that would otherwise need to be solved numerically using a boundary-element method for a two-dimensional crack in a three-dimensional medium. Furthermore, the asymptotic solutions presented here provide closed-form expressions of the solutions essentially over the entire range of $T$, as evidenced, for example in Figures 1 and 6. The utility of an analytical, reduced-order model, for example, is apparent when considering it as an elementary model for the inference of aseismic slip due to a known or suspected, natural or artificial, fluid source that drives aseismic slip. Not having to run an elaborate forward numerical model at each iteration of an inversion drastically reduces computation time. Such an axisymmetric representation was used, for example, by \emph{Bhattacharya and Viesca} [2019] in the inversion of hydromechanical properties for a time-variable injection history. A constant injection rate history is a reasonable starting point for a case in which data are too poor to constrain an injection history, as would likely be the case for any naturally occurring aseismic rupture in an interplate or crustal fault setting, such as, for example, the observed concentric migration of microseismicity thought to be due to fluid migration [\emph{Ross et al.}, 2020]. Here, for an assumed set of properties of the fault and surrounding rock (friction coefficient, hydraulic diffusivity, shear modulus) and values of the initial fault tractions and quantities related to injection (as they enter lumped sets of parameters determining $\Delta p$ and $T$), the aseismic source time function, or moment history $M(t)$, which enter the far-field representations of quasi-static displacement and stress fields used for comparison with observations, is directly determined by the presented solutions, eliminating the need for a forward-model solution step.

\

The additional utility of such asymptotic expansions is as a benchmark for numerical methods for capable of solving the posed problem and as a basis for further analytical work. The problem explored here is the first instance of a crack in a three-dimensional medium (fluid-driven or otherwise) having a perturbation solution carried to arbitrary order. \emph{S\'aez et al.} [2023] were able to used an incomplete leading-order solution to this end, missing only the inner, boundary-layer solution in the critically stressed limit. With the complete perturbation expansions, precise determination of the absolute error of numerical procedures can be attained. This contribution is timely as long-standing initiatives for benchmarking models of sequences of earthquakes and seismic slip (SEAS) [e.g., \emph{Harris et al.}, 2009; \emph{Erickson et al.}, 2020] have recently turned their attention towards models of fluid-induced aseismic slip [\emph{Lambert et al.}, 2025]. This initiative has incorporated earlier results for simpler, two-dimensional problems [\emph{Viesca}, 2021] in benchmark determination. The methods used here and in previous work [\emph{Viesca}, 2021; \emph{S\'aez et al}, 2023], have motivated the development of analytical results accounting for more realistic descriptions of fault behavior, including slip-driven enhancement of the along-fault permeability and fault dilatancy [\emph{Dunham}, 2024]. Additionally, recent results also show that such relatively simple models can effectively represent the behavior, including slip distribution and rupture extent, of models with more elaborate, rate- and state-dependent models for friction coefficient and permeability evolution [\emph{Berrios-Rivera et al.}, 2025].
\section{Conclusion}
In this article, we provided asymptotic perturbation expansions to arbitrary order for a problem of an expanding, mixed-mode circular shear fracture.
We demonstrated that expressing these solutions to arbitrary order allows for accurate representation of the full solution over the entire range of the problem parameter, including when the parameter is neither large nor small. 
The specific problem pursued here provided the opportunity to illustrate both regular and singular perturbations methods as applied to singular integral equations  intrinsic to fracture mechanics, with the singular perturbation involving a composite solution consisting of a boundary layer asymptotically matched with an outer solution.
We detailed the derivation of these solutions, and in the case of the outer solution we provide an alternative derivation by multipole expansion.

\

\backsection[Funding]{This work was supported by the National Science Foundation (grant number 1653382).}

\backsection[Declaration of interests]{The authors report no conflict of interest.}

\backsection[Author ORCID]{R. C. Viesca, https://orcid.org/0000-0003-4180-7807}

\appendix

\section{Asymptotic solution details of inner solution for critically stressed conditions}
\label{app:in}
Here we derive the boundary-layer solution for slip near $r=0$ valid over distances comparable to the diffusion lengthscale $\sqrt{\alpha t}$. However, we note that in our original inversion for slip (\ref{eq:dt}) and (\ref{eq:dtf}), radial distances are scaled by the current rupture radius $R$. Here we revisit the problem of solving for slip, but now scale radial distances by  $\sqrt{\alpha t}$, denoting
$$s=r/\sqrt{\alpha t}$$
Substituting the variable of integration in (\ref{eq:dtf}) with $\eta=\xi/\lambda$, we may thus rewrite (\ref{eq:dtf}) as 
\begin{equation}
\delta(s,t)=-\frac{4}{\pi}\frac{\sqrt{\alpha t}}{\mu}\int_s^\lambda  \sqrt{\xi^2-s^2}\frac{df}{d\xi}d\xi
\label{eq:ds}
\end{equation}
where
\begin{equation}
\frac{df}{d\xi}=\frac{d}{d\xi}\int_0^1 \frac{x \Delta\tau(x\xi \sqrt{\alpha t})}{\sqrt{1-x^2}}dx
\label{eq:dfdxi}
\end{equation}
in which we have suppressed explicit time dependence of $\Delta\tau$, except through the term $\sqrt{\alpha t}$. That we may do so is apparent considering that the 
 particular expression for $\Delta\tau$ of interest follows the combination of (\ref{eq:dtau}) and (\ref{eq:f1})--\ref{eq:R}) and the requirement that $\tau=\tau_s$,
$$\Delta\tau(r,t)=\tau_b-f\left[\sigma'-\Delta p E_1(r^2/\alpha t)\right]$$
evaluating
$$\Delta\tau(x\xi \sqrt{\alpha t},t)=\tau_b-f\left(\sigma'-\Delta p E_1[(x\xi)^2]\right)$$ 
In this case, we can simplify $\Delta\tau(x\xi \sqrt{\alpha t},t)\Rightarrow\Delta\tau(x\xi)$ and do so in our evaluation of (\ref{eq:dfdxi}),
\begin{align*}
\frac{df}{d\xi}&=\frac{d}{d\xi}\int_0^1 \frac{x \Delta\tau(x\xi)}{\sqrt{1-x^2}}dx\\[9 pt]
&=\int_0^1\frac{x}{\sqrt{1-x^2}} \displaystyle\frac{d\Delta\tau}{dw}{\displaystyle\frac{dw}{d\xi}}dx\\[9 pt]
&=-2f\Delta p \, \frac{1}{\xi}\int_0^1 \frac{x e^{-(x\xi)^2}}{\sqrt{1-x^2}}dx
\end{align*}
where in the above $w=x \xi$ and $d\Delta\tau(w)/dw=2 \exp[-(x \xi)^2]/(x\xi)$. We now label the integral
\begin{equation}F(\xi)=\int_0^1 \frac{x e^{-(x\xi)^2}}{\sqrt{1-x^2}}dx\label{eq:Fxi}\end{equation}
such that (\ref{eq:ds}) becomes
$$\delta(s,t)=\frac{\sqrt{\alpha t}}{\mu} f \Delta p\frac{8}{\pi} \int_s^\lambda\sqrt{\xi^2-s^2}\frac{F(\xi)}{\xi}d\xi$$
or more simply
\begin{equation}
\frac{\delta(s,t)}{\sqrt{\alpha t} f \Delta p/\mu}=\frac{8}{\pi} \int_s^\lambda \sqrt{1-(s/\xi)^2}\,F(\xi)\,d\xi
\label{eq:da}
\end{equation}

We now look to asymptotically evaluate the expression (\ref{eq:da}), which is exact, in the limit of large $\lambda$ to arrive to our expansion approximation of the inner solution for critically stressed conditions. To do so, we will first rewrite (\ref{eq:da}) as:
\begin{equation}
\frac{\delta(s,t)}{\sqrt{\alpha t} f \Delta p/\mu}=\frac{8}{\pi}\left[\int_s^\infty \sqrt{1-(s/\xi)^2}F(\xi)d\xi-\int_\lambda^\infty \sqrt{1-(s/\xi)^2}F(\xi)d\xi\right]
\label{eq:dar}
\end{equation}
where the leading-order term, $O(1)$, is 
\begin{equation}\frac{8}{\pi}\int_s^\infty \sqrt{1-(s/\xi)^2}F(\xi)d\xi=2\sqrt{\pi}e^{-s^2/2} \left[(1+s^2)I_0(s^2/2)+s^2 I_1(s^2/2)\right]-4s\label{eq:d0OG}\end{equation}
where $I_0$ and $I_1$ modified Bessel functions of the first kind.
We turn our attention to the second integral in (\ref{eq:dar}) and note that, for large $\lambda$, the integration variable $\xi$ takes on only large values warranting a Taylor expansion of the integrand's square-root term
\begin{equation}-\int_\lambda^\infty \sqrt{1-(s/\xi)^2}F(\xi)d\xi\approx -\int_\lambda^\infty \left(1-\frac{1}{2}(s/\xi)^2-\frac{1}{8}(s/\xi)^4+...\right) F(\xi)d\xi\label{eq:dinF}\end{equation}
We look to evaluate the series of integrals of (\ref{eq:dinF}), which are the remaining $O(1/\lambda)$ terms of (\ref{eq:dar}). 

\

We start with the first integral of the expansion (\ref{eq:dinF}),
$$I(\lambda)=-\int_\lambda^\infty F(\xi)d\xi$$
To expand $I$ for large $\lambda$, we note that 
$$\frac{dI}{d\lambda}=F(\lambda)$$
and seek the expansion of
\begin{equation}F(\lambda)=\int_0^1 \frac{x e^{-(\lambda x)^2}}{\sqrt{1-x^2}}dx\label{eq:F}\end{equation}
This will be useful not only to arrive to $I(\lambda)$, but also to the remaining integrals in the expansion (\ref{eq:dinF}). We use Laplace's method for asymptotic expansion, substituting the binomial-like expansion for the Maclaurin series of $(1-x^2)^{-1/2}$
\begin{equation}
\frac{1}{\sqrt{1-x^2}}=\sum_{k=0}^\infty \begin{pmatrix}
-1/2\\
k
\end{pmatrix} (-x^2)^k=\sum_{k=0}^\infty \frac{(2k)!}{(k!)^2}\left(\frac{x}{2}\right)^{2k}
\label{eq:mac}
\end{equation}
and passing the upper limit of integration to $\infty$ without loss of generality for the power-law asymptotic expansion
\begin{equation*}F(\lambda)\sim\sum_{k=0}^\infty \frac{(2k)!}{(k!)^2} \frac{1}{2^{2k}} \int_0^\infty x^{2k+1} e^{-(\lambda x)^2}dx\end{equation*}
we may evaluate the integrals using the change of variable $u=(\lambda x)^2$
\begin{equation*}
\int_0^\infty x^{2k+1} e^{-(\lambda x)^2}dx=\frac{1}{2\lambda} \int_0^\infty \left(\frac{\sqrt{u}}{\lambda}\right)^{2k+1} e^{-u}du=\frac{\Gamma(k+1)}{2 \lambda^{2k+2}}=\frac{k!}{2 \lambda^{2k+2}}
\end{equation*}
such that
\begin{equation*}F(\lambda)\sim\sum_{k=0}^\infty \frac{1}{\lambda^{2k+2} }\frac{1}{2^{2k+1}} \frac{(2k)!}{k!}\end{equation*}
Since $(2k)!=(2k+1)!/(2k+1)$ and $n!=\Gamma(n+1)$, we may rewrite the above as 
\begin{equation*}F(\lambda)\sim\sum_{k=0}^\infty \frac{1}{\lambda^{2k+2} }\frac{1}{2^{2k+1}} \frac{\Gamma(2k+2)}{\Gamma(k+1)}\frac{1}{2k+1}\end{equation*}
Using Legendre's duplication formula for the Gamma function
\begin{equation}2^{1-m}\,\Gamma(m)/\Gamma(m/2)=\Gamma(m/2+1/2)/\sqrt{\pi}\label{eq:Leg}\end{equation}
while identifying $m=2k+2$, we may simplify the expression for $F(\lambda)$ to
\begin{equation}F(\lambda)\sim\sum_{k=0}^\infty \frac{1}{\sqrt{\pi}} \frac{1}{\lambda^{2k+2} }\frac{\Gamma(k+3/2)}{2k+1}\label{eq:F3}\end{equation}
and we may finally asymptotically approximate the integral
$$I(\lambda)=-\int_\lambda^\infty F(\lambda)d\lambda\sim-\frac{1}{\sqrt{\pi}}\sum_{n=0}^\infty \frac{1}{\lambda^{2n+1}} \frac{\Gamma(n+3/2)}{(2n+1)^2}$$

\

We now look to evaluate all terms in the expansion (\ref{eq:dinF}), which can be written in summation form as
$$-\int_\lambda^\infty \sqrt{1-(s/\xi)^2}F(\xi)d\xi=-\int_\lambda^\infty \left[-\sum_{k=0}^\infty \frac{(k-3/2)!}{2\sqrt{\pi} k!}\left(\frac{s}{\xi}\right)^{2k}\right] F(\xi)d\xi$$
where the summation in brackets is the Taylor series of $\sqrt{1-(s/\xi)^2}$ about $s/\xi=0$. Again passing from factorials to Gamma functions, via the identity $\Gamma(n+1)=n!$, and rearranging we have 
\begin{equation}-\int_\lambda^\infty \sqrt{1-(s/\xi)^2}F(\xi)d\xi= \sum_{k=0}^\infty \frac{1}{2\sqrt{\pi}}\frac{\Gamma(k-1/2)}{\Gamma(k+1)}s^{2k} \int_\lambda^\infty \frac{F(\xi)}{\xi^{2k}} d\xi \label{eq:J}\end{equation}
we denote the integral
$$J(\lambda)=\int_\lambda^\infty \frac{F(\xi)}{\xi^{2k}} d\xi $$ such that
$$\frac{dJ}{d\lambda}=-\frac{F(\lambda)}{\lambda^{2k}}\sim-\frac{1}{\sqrt{\pi}}\frac{1}{\lambda^{2k}}\left(\sum_{m=1,3,...}^\infty \frac{1}{\lambda^{2n+2}}\frac{\Gamma(n+3/2)}{2n+1}\right)$$
in which we have substituted the previously found asymptotic expansion for $F(\lambda)$, (\ref{eq:F3}). We may subsequently integrate to find
$$J(\lambda)=-\int_\lambda^\infty\frac{dJ}{d\lambda}d\lambda\sim\frac{1}{\sqrt{\pi}}\sum_{n=0}^\infty \frac{1}{\lambda^{2(n+k)+1}}\frac{1}{2(n+k)+1}\frac{\Gamma(n+3/2)}{2n+1}$$
Substituting this expression for $J$ into (\ref{eq:J})
\begin{align}-\int_\lambda^\infty \sqrt{1-(s/\xi)^2}F(\xi)d\xi&\sim\frac{1}{2\pi}\sum_{k=0}^\infty\sum_{n=0}^\infty s^{2k}\frac{\Gamma(k-1/2)}{\Gamma(k+1)}\frac{1}{\lambda^{2(n+k)+1}}\frac{1}{2(n+k)+1}\frac{\Gamma(n+3/2)}{2n+1}\label{eq:a}\\[9 pt]
&\sim\frac{1}{2\pi}\sum_{k=0}^\infty\sum_{j=k}^\infty s^{2k}\frac{\Gamma(k-1/2)}{\Gamma(k+1)}\frac{1}{\lambda^{2j+1}}\frac{1}{2j+1}\frac{\Gamma(3/2+j-k)}{2(j-k)+1}\label{eq:b}\\[9 pt]
&\sim\frac{1}{2\pi}\sum_{j=0}^\infty\frac{1}{\lambda^{2j+1}}\frac{1}{2j+1}\left[\sum_{k=0}^j s^{2k}\frac{\Gamma(k-1/2)}{\Gamma(k+1)}\frac{\Gamma(3/2+j-k)}{2(j-k)+1}\right]\label{eq:c}\end{align}
where, in passing from (\ref{eq:a}) to (\ref{eq:b}) we replaced the summation index $n$ with another $j$ chosen such that $2j+1=2(n+k)+1$, or $n=j-k$, and in passing from (\ref{eq:b}) to (\ref{eq:c}) we swapped the order of summation. We thus find that the remaining term in (\ref{eq:dar}) is asymptotically expanded for large $\lambda$ as
\begin{equation}
\frac{8}{\pi}\left(-\int_\lambda^\infty \sqrt{1-(s/\xi)^2}F(\xi)d\xi \right)\sim -\frac{1}{\lambda}\frac{4}{\pi}-\frac{1}{\lambda^3}\frac{2}{3\pi}(1-s^2)-\frac{1}{\lambda^5}\frac{1}{5\pi}\left(3-s^2-s^4/2\right)+O(\lambda^{-7})
\end{equation}
We therefore arrive to the asymptotic expansion of the inner solution for slip in the critically stressed limit, $\delta\sim\delta_{in}$ as $\lambda\rightarrow\infty$ where
\begin{equation}
\frac{\delta_{in}(s,t)}{\sqrt{\alpha t} f\Delta p/\mu}=\delta_0(s,t)-\frac{1}{\lambda}\frac{4}{\pi}-\frac{1}{\lambda^3}\frac{2}{3\pi}(1-s^2)-\frac{1}{\lambda^5}\frac{1}{5\pi}\left(3-s^2-s^4/2\right)+O(\lambda^{-7})
\end{equation}
and
$$\delta_0(s,t)=2\sqrt{\pi}e^{-s^2/2} \left[(1+s)I_0(s^2/2)+s^2 I_1(s^2/2\right]-4s$$
the outer behavior of which as $s\rightarrow\infty$ is
\begin{equation}
\frac{\delta_{in}(s\rightarrow\infty,t)}{\sqrt{\alpha t} f\Delta p/\mu}=\left(\frac{1}{s}+\frac{1}{8 s^3}+\frac{3}{32 s^5}+O(s^{-7})\right)-\frac{1}{\lambda}\frac{4}{\pi}-\frac{1}{\lambda^3}\frac{2}{3\pi}(1-s^2)-\frac{1}{\lambda^5}\frac{1}{5\pi}\left(3-s^2-s^4/2\right)+O(\lambda^{-7})
\end{equation}

\section{Asymptotic solution details of outer solution for critically stressed conditions}
\label{app:out}
Here we turn our attention to the solution for slip for radial distances $r$ comparable to the crack-tip radius $R(t)$ when $R(t)\gg \sqrt{\alpha t}$, i.e., when $\lambda$ is large. We return to the inversion for slip (\ref{eq:dtfbp}), applying the non-singular condition $f(1)=0$: 
\begin{equation}
\delta(\rho,t)=-\frac{4}{\pi}\frac{R}{\mu}\int_\rho^1 \sqrt{\eta^2-\rho^2}\frac{d}{d\eta}\int_0^1 \frac{x \Delta\tau(x\eta R,t)}{\sqrt{1-x^2}}dxd\eta
\label{eq:dfdeta}
\end{equation}
where we recall that
$$\Delta \tau(r,t)=\tau_b-f[\sigma'-\Delta p E_1(r^2/\alpha t)]$$
When evaluating
$$\Delta\tau(x\eta R(t),t)=\tau_b-f\left(\sigma'-\Delta p E_1[(x\eta\lambda)^2]\right)$$
we note that we can simplify the left-hand side $\Delta\tau(x\xi \sqrt{\alpha t},t)\Rightarrow\Delta\tau(x\eta\lambda)$ and do so in our evaluation of (\ref{eq:dfdeta}), as was done similarly for the inner solution in Appendix A:
\begin{align*}
\frac{d}{d\eta}\int_0^1 \frac{x \Delta\tau(x\eta \lambda)}{\sqrt{1-x^2}}dx&=\int_0^1 \frac{x}{\sqrt{1-x^2}}\frac{d\Delta\tau}{dw}\frac{dw}{d\zeta}dx\\[9 pt]
&=-2f\Delta p \frac{1}{\eta}\int_0^1\frac{xe^{-(x\eta\lambda)^2}}{\sqrt{1-x^2}}dx
\end{align*}
where now $w=x\eta \lambda$, and (\ref{eq:dfdeta}) becomes
\begin{equation}
\delta(\rho,t)=\frac{R f \Delta p}{\mu}\frac{8}{\pi}\int_\rho^1 \frac{\sqrt{\eta^2-\rho^2}}{\eta}\int_0^1 \frac{x e^{-(x\eta\lambda)^2}}{\sqrt{1-x^2}}dx d \eta
\label{eq:dRF}
\end{equation}
We identify the innermost integral, following (\ref{eq:F}), as $F(\eta\lambda)$ which has the known asymptotic expansion for large values of its argument (\ref{eq:F3}). We substitute this expansion in (\ref{eq:dRF}) to find the asymptotic expansion of the outer solution for slip, $\delta\sim\delta_{out}$ where
\begin{equation}
\delta_{out}(\rho,t)=\frac{R f \Delta p}{\mu}\frac{8}{\pi} \sum_{n=0}^\infty \frac{1}{\lambda^{2n+2}} \frac{1}{\sqrt{\pi}}\frac{\Gamma(n+3/2)}{2n+1} \int_\rho ^1 \frac{\sqrt{\eta^2-\rho^2}}{\eta^{2n+3}}d\eta
\label{eq:docomp}
\end{equation}
To evaluate the integral, we use the trigonometric variable substitution $\cos\theta=\rho/\eta$ on 
$$\int_\rho ^1 \frac{\sqrt{\eta^2-\rho^2}}{\eta^{2n+3}}d\eta=\frac{1}{\rho^{2n+1}}\int_0^{\text{acos}(\rho)}\cos^{2n}\theta \sin^2\theta d\theta$$
and to evaluate this last integral in turn, we will consider the general integral
\begin{equation} I(m,p,\rho)=\int_0^{\text{acos}(\rho)}\cos^{m}\theta \sin^p\theta\, d\theta \label{eq:Impr}\end{equation}
whose solution is as follows. We begin by rearranging the integrand 
\begin{align*}
I(m,p,\rho)&=\int_0^{\text{acos}(\rho)}\left(\cos^{m-1}\theta\right)\left( \sin^p\theta \cos\theta\right)d\theta\\[9 pt]
&=\frac{\rho^{m-1}(1-\rho^2)^{\frac{p+1}{2}}}{p+1}+\frac{m-1}{p-1}\int_0^{\text{acos}(\rho)}\cos^{m-2}\theta  \sin^{p+2}d\theta\\[9 pt]
&=\frac{\rho^{m-1}(1-\rho^2)^{\frac{p+1}{2}}}{p+1}-\frac{m-1}{p+1}\int_0^{\text{acos}(\rho)}\cos^m\theta\sin^p\theta d\theta+\frac{m-1}{p+1}\int_0^{\text{acos}(\rho)}\cos^{m-2}\theta\sin^p\theta d\theta
\end{align*}
and, in the above, we pass from the first to second line using integration by parts, and from the second to third via $\sin^2\theta=1-\cos^2\theta$. We identify the first integral as $I(m,p,\rho)$ and the second as $I(m-2,p,\rho)$. Solving for $I(m,p,\rho)$, we find the recursion relation:
\begin{equation}I(m,p,\rho)=\frac{\rho^{m-1}(1-\rho^2)^{\frac{p+1}{2}}}{m+p}+\frac{m-1}{m+p}I(m-2,p,\rho)\label{eq:Imp}\end{equation}
Returning to the specific case $m=2n$ and $p=2$, we find the recursion relation for which we need only calculate the case $n=0$ to find all others $n>0$:
\begin{align*}
I(0,2,\rho)&=\frac{1}{2}\left(\text{acos}(\rho)-\rho\sqrt{1-\rho^2}\right)\\[9 pt]
I(2,2,\rho)&=\frac{1}{8}\left(\text{acos}(\rho)-\rho\sqrt{1-\rho^2}\right)+\frac{\rho(1-\rho^2)^{3/2}}{4}\\[9 pt]
&=\frac{1}{8}\left[\left(\text{acos}(\rho)-\rho\sqrt{1-\rho^2}\right)+\frac{(1-\rho^2)^{3/2}}{\rho} 2\rho^2\right]\\[9 pt]
I(4,2,\rho)&=\frac{1}{16}\left(\text{acos}(\rho)-\rho\sqrt{1-\rho^2}\right)+\frac{\rho(1-\rho^2)^{3/2}}{8}+\frac{\rho^3(1-\rho^2)^{3/2}}{6}\\[9pt]
&=\frac{1}{16}\left[\left(\text{acos}(\rho)-\rho\sqrt{1-\rho^2}\right)+\frac{(1-\rho^2)^{3/2}}{\rho}\left(2\rho^2+\frac{8}{3}\rho^4\right)\right]
\end{align*}
From this recursion, we may deduce an explicit expression for $I(m,2,\rho)$, recognizing from the sequence above that such an expression will have the form
$$I(m,2,\rho)= a_m\left[\left(\text{acos}(\rho)-\rho\sqrt{1-\rho^2}\right)+\frac{(1-\rho^2)^{3/2}}{\rho}\sum_{k=2,4,...}^mb_k\rho^k\right]$$
where $a_m$ and $b_k$ are to be determined. From the factor $(m-1)/(m+2)$ in the recursion, we recognize that
\begin{align}
a_m=\frac{(m-1)}{(m+2)}\cdot\frac{(m-3)}{m}\cdot\frac{(m-5)}{(m-2)}\cdot...\cdot\frac{7}{10}\cdot \frac{5}{8}\cdot \frac{3}{6}\cdot \frac{1}{4} \cdot \frac{1}{2}
\end{align}
Substituting $m=2n$
\begin{align}
a_{2n}&=\frac{(2n-1)(2n-3)(2n-5)...}{(2n+2)(2n)(2n-2)...}\\[9 pt]
&=\frac{(2n-1)(2n-3)(2n-5)...}{2^{n+1}(n+1)(n)(n-1)...} \\[9 pt]
&=\frac{(2n-1)!/[2^{n-1}(n-1)!]}{2^{n+1}(n+1)!}\\[9 pt]
&=\frac{(2n-1)!}{2^{2n}(n+1)!(n-1)!}
\end{align}
Passing to the gamma function $n!=\Gamma(n+1)$
$$a_{2n} = \frac{\Gamma(2n)}{2^{2n}\Gamma(n)\Gamma(n+2)}$$
Using the Legendre duplication identity $\Gamma(2z)/[2^{2z}\Gamma(z)]=\Gamma(z+1/2)/(2\sqrt{\pi})$
$$ a_{2n}=\frac{1}{2\sqrt{\pi}}\frac{\Gamma(n+1/2)}{\Gamma(n+2)}$$
$$a_m=\frac{1}{2\sqrt{\pi}}\frac{\Gamma\left(\frac{m}{2}+\frac{1}{2}\right)}{\Gamma\left(\frac{m}{2}+2\right)}$$
Regarding the coefficients of the series $b_k$, we can immediately deduce that $b_0=0$, and may find from the recursion (\ref{eq:Imp}) and the factoring of $a_m$ that, for $k\geq 2$
$$b_k=\frac{1}{a_k} \frac{1}{k+2}$$

\

We may now write the asymptotic expansion for the outer solution under critically stressed conditions:
\begin{equation*}
\frac{\delta_{out}(\rho,t)}{R f \Delta p/\mu}=\frac{8}{\pi}\sum_{n=0}^\infty\frac{1}{\lambda^{2n+2}}\frac{1}{\sqrt{\pi}}\frac{\Gamma(n+3/2)}{2n+1}\frac{1}{\rho^{2n+1}}a_{2n}\left[\left(\text{acos}(\rho)-\rho\sqrt{1-\rho^2}\right)+\frac{(1-\rho^2)^{3/2}}{\rho}\sum_{k=1}^{n} b_{2k}\rho^{2k}\right]
\end{equation*}
\begin{align}\begin{split}
\frac{\delta_{out}(\rho,t)}{R f \Delta p/\mu}=\frac{4}{\pi^{2}}\sum_{n=0}^\infty&\frac{1}{\lambda^{2n+2}}\frac{1}{\rho^{2n+1}}\frac{\Gamma(n+3/2)}{2n+1}\frac{\Gamma(n+1/2)}{\Gamma(n+2)}  \\[9 pt]
&\cdot\left[\left(\text{acos}(\rho)-\rho\sqrt{1-\rho^2}\right)+2\sqrt{\pi}\frac{(1-\rho^2)^{3/2}}{\rho}\sum_{k=1}^{n}\frac{\rho^{2k}}{2k+2} \frac{\Gamma(k+2)}{\Gamma(k+1/2)}\right]
\label{eq:doutapp}
\end{split}\end{align}

\section{Alternative derivation of critically stressed outer solution using multipole expansion}
\label{app:outa}

We begin by using the inverted solution for slip in terms of the change in shear stress within the crack, repeated below
\begin{equation}
\delta(\rho,t)=-\frac{4}{\pi}\frac{R}{\mu}\int_\rho^1 \sqrt{\eta^2-\rho^2}\frac{d}{d\eta}\int_0^1 \frac{x \Delta\tau(x\eta R,t)}{\sqrt{1-x^2}}dxd\eta
\label{eq:apd}
\end{equation}
where we recall that the change in shear stress from the background level $\Delta\tau(r,t)=\tau_b-\tau(r,t)$ and that we require that the shear stress equals the shear strength $\tau_s=f[\sigma -p(r,t)]$, with the fluid pressure distribution being
\begin{equation}
p(r,t)=p_o+\Delta pE_1[r^2/(\alpha t)]
\label{eq:ap}
\end{equation}

In the critically stressed regime, in which the fluid pressure distribution appears concentrated about the crack origin if viewed over a lengthscale comparable to the crack size, we approximate the fluid pressure distribution using a multipole expansion
\begin{equation}
p(r,t)=p_o+\Delta p \left[\frac{p_0(t)}{2\pi r}\delta_D(r)-\frac{p_1(t)}{2\pi r}\delta_D'(r)+\frac{p_2(t)}{2\pi r}\delta_D''(r)-...+\frac{p_n(t)}{2\pi r}\delta_D^{(n)}(r)\right]
\label{eq:pm}
\end{equation}
where $\delta_D(r)$ is the Dirac delta, $\delta^{(n)}_D(r)$ denotes its $n$-th derivative, and $p_n(t)$ is the $n$-th moment of the distribution of the scaled, axisymmetric pressure change
\begin{equation}p_n(t)=\int_0^\infty \frac{r^n}{n!} \frac{p(r,t)-p_o}{\Delta p} 2\pi r dr\end{equation}
which, for the change due to injection (\ref{eq:ap}), evaluates to
\begin{equation}
p_n(t)= (\alpha t)^{1+n/2}\,\frac{\Gamma(1+n/2)}{(1+n/2)}\frac{\pi}{n!}
\label{eq:pn}
\end{equation}
While $p_n$ will generally be non-zero, we will find that only the even terms in the multipole expansion of $p(r,t)$ will have non-zero contributions to the slip profile. We also recall that the Dirac delta has the property
\begin{equation}\int_0^\infty \delta^{(n)}(r a)f(r)dr=\frac{(-1)^n}{a^{n+1}}f^{(n)}(0)\label{eq:dd}\end{equation}

\

Expanding the expression (\ref{eq:apd}),
\begin{equation}\delta(\rho,t)= -\frac{4}{\pi}\frac{R}{\mu} \int_\rho^1 \sqrt{\eta^2-\rho^2}\frac{d}{d\eta}\int_0^1\frac{x \left[\tau_b-f(\sigma-p(x\eta R,t))\right] }{\sqrt{1-x^2}}dx\,d\eta\end{equation}
and rearranging
\begin{equation}\delta(\rho,t)= -\frac{4}{\pi}\frac{R f\Delta p}{\mu} \int_\rho^1\sqrt{\eta^2-\rho^2}\frac{d}{d\eta}\int_0^1\frac{x}{\sqrt{1-x^2}} \left[\frac{\tau_b-f\sigma'}{f\Delta p} + \frac{p(x\eta R,t)-p_o}{\Delta p}\right] dx\,d\eta\end{equation}
Considering that a derivative with respect to $\eta$ operates on the innermost integral, we may drop the first term in brackets (identically $T$) as it is independent of $\eta$. Additionally, we can introduce the multipole expansion for the second term of (\ref{eq:pm}) in brackets to calculate the outer solution  asymptotically approximating slip, $\delta\sim\delta_{out}$
\begin{equation}\delta_{out}(\rho,t)= -\frac{4}{\pi}\frac{R f\Delta p}{\mu} \left(\frac{1}{2\pi}\int_\rho^1\sqrt{\eta^2-\rho^2}\frac{d}{d\eta}\left[\frac{1}{\eta R}\int_0^1\frac{1}{\sqrt{1-x^2}} \sum_{n=0}^\infty (-1)^n p_n(t)\delta_D^{(n)}(x\eta R) dx\right]d\eta\right)\end{equation}
We may use the property of the Dirac delta (\ref{eq:dd})  to evaluate the innermost integral
\begin{equation}\delta_{out}(\rho,t)= \frac{4}{\pi}\frac{R f\Delta p}{\mu} \left(\frac{1}{2\pi}\int_\rho^1\sqrt{\eta^2-\rho^2}\frac{d}{d\eta}\left[\frac{1}{\eta R}\sum_{n=0}^\infty \frac{p_n(t)}{(\eta R)^{n+1}} \frac{d}{dx^n}\left.\left(\frac{1}{\sqrt{1-x^2}}\right)\right |_{x=0}\right]d\eta\right)\end{equation}
We can immediately recognize that odd $n$-th derivatives of the even function $1/\sqrt{1-x^2}$ evaluated at the origin will be zero, due to symmetry. Thus all odd terms in the sum have no contribution. Rather than continuing the sum as $n=0,2,4...$, we pass the summation index $n\rightarrow 2n$ with the new index maintaining the progression $n=0,1,2,...$ . From the Maclaurin series (\ref{eq:mac}) we find that, for this new index $n$,
$$\frac{d}{dx^n}\left.\left(\frac{1}{\sqrt{1-x^2}}\right)\right |_{x=0}=\left(\frac{(2n)!}{2^{n} \,n !}\right)^2 $$
such that
\begin{equation}\delta_{out}(\rho,t)= \frac{4}{\pi}\frac{R f\Delta p}{\mu} \left(\frac{1}{2\pi}\int_\rho^1\sqrt{\eta^2-\rho^2}\frac{d}{d\eta}\left[\frac{1}{\eta R}\sum_{n=0}^\infty \frac{p_{2n}(t)}{(\eta R)^{2n+1}} \left(\frac{(2n)!}{2^{n} \,n!}\right)^2\right]d\eta\right)\end{equation}
Evaluating the derivative we arrive to 
\begin{equation}\delta_{out}(\rho,t)= \frac{4}{\pi}\frac{R f\Delta p}{\mu} \left(\frac{1}{2\pi}\sum_{n=0}^\infty  \frac{p_{2n}(t)}{R^{2n+2}}\left(\frac{(2n)!}{2^{n}\,n!}\right)^2 (2n+2)\int_\rho^1\frac{\sqrt{\eta^2-\rho^2}}{\eta^{2n+3}}\,d\eta\right)\end{equation}
Using (\ref{eq:pn}) to evaluate $p_{2n}$ and recognizing that $(\alpha t)^{1+n}/R^{2n +2}=1/\lambda^{2n+2}$, this further simplifies to
\begin{equation}\delta_{out}(\rho,t)= \frac{4}{\pi}\frac{R f\Delta p}{\mu} \sum_{n=0}^\infty  \frac{1}{\lambda^{2n+2}}\frac{(2n)!}{2^{2n}\,n!} \int_\rho^1\frac{\sqrt{\eta^2-\rho^2}}{\eta^{2n+3}}\,d\eta\end{equation}
with which we may compare the solution for $\delta_{out}$ found by direct asymptotic expansion, (\ref{eq:docomp}). The equivalence of these two expressions for $\delta_{out}$ follows from the equivalence of the numerical coefficients
$$\frac{8}{\pi} \frac{1}{\sqrt{\pi}}\frac{\Gamma(n+3/2)}{2n+1}=\frac{4}{\pi}\frac{(2n)!}{2^{2n} n!}$$
which directly follows from the Legendre-duplication formula, (\ref{eq:Leg}). The final expression for the outer solution can then be found by repeating the steps subsequent to (\ref{eq:docomp}).

\section{Asymptotic solution details for marginally pressurized conditions}
\label{app:mp}
We pursue the regular perturbation expansion of the solution in the marginally pressurized limit, $\lambda\ll1$. We return to the integral expression for slip due to the constant-injection-rate source
$$\delta(\rho,t)=\frac{8}{\pi}\frac{R f\Delta p}{\mu}\int_\rho^1 \frac{\sqrt{\eta^2-\rho^2}}{\eta}\int_0^1 \frac{x e^{-(x\eta\lambda)^2}}{\sqrt{1-x^2}}dxd\eta$$
where we again identify the innermost integral as $F(\eta\lambda)$ and look to expand it now for $\eta\lambda\ll1$. In the above, we can substitute the Taylor expansion
$$e^{-(x\eta\lambda)^2}=\sum_{n=0}^\infty \frac{(-1)^{n}}{n!} (x\eta \lambda)^{2n}$$
and subsequently evaluate integrals of the form
$$ I_{2n+1}=\int_0^1 \frac{x^{2n+1}}{\sqrt{1-x^2}}dx$$
for $n=0,1,2...$
Letting $m=2n+1$ and $x=\sin\theta$ we transform the integral to
\begin{align*} I_m&=\int_0^1 \frac{x^m}{\sqrt{1-x^2}}dx=\int_0^{\pi/2}\sin^m\theta d\theta \\[9 pt]
&=\int_0^{\pi/2}\sin^{m-1}\theta \sin\theta d\theta=\left(-\sin^{m-1}\theta\cos\theta\right | _0^{\pi/2}+(m-1)\int_0^{\pi/2}\sin^{m-2}\theta\cos^2\theta d\theta \\[9 pt]
&=(m-1)\left(\int_0^{\pi/2}\sin^{m-2}\theta d\theta-\int_0^{\pi/2}\sin^{m}\theta d \theta\right)\\[9 pt]
&=(m-1)(I_{m-2}-I_m)
\end{align*}
and arrive to the recursion relation
$$I_m=\frac{m-1}{m}I_{m-2}$$
Given that $I_1=1$,
\begin{equation}I_{2n+1}=\frac{(2n)\cdot(2n-2)\cdot...\cdot6\cdot4\cdot 2}{(2n+1)\cdot(2n-1)\cdot...\cdot7\cdot 5\cdot 3}= \frac{2^n n!}{(2n+1)!/(2^n n!)}\label{eq:seq}\end{equation}
which we may rewrite using the Gamma function
$$I_{2n+1}=4^n \frac{\left[\Gamma(n+1)\right]^2}{\Gamma(2n+2)}$$
and using the Legendre duplication formula we can further simplify
$$I_{2n+1}=\frac{\sqrt{\pi}}{2} \frac{\Gamma(n+1)}{\Gamma(n+3/2)}$$
and arrive to an expression for the expansion of $F$ for small values of its argument
$$F(\eta\lambda)=\frac{\sqrt{\pi}}{2}\sum_{n=0}^\infty (\eta\lambda)^{2n} \frac{(-1)^n}{\Gamma(n+3/2)}$$

$$\frac{\delta(\rho,t)}{R f \Delta p/\mu}=\frac{8}{\pi}\int_\rho ^1 \frac{\sqrt{\eta^2-\rho^2}}{\eta} F(\eta\lambda)d\eta=\frac{4}{\sqrt{\pi}}\sum_{n=0}^\infty \frac{(-1)^n}{\Gamma(n+3/2)}\lambda^{2n} \int_\rho^1 \sqrt{\eta^2-\rho^2}\eta^{2n-1}d\eta $$
Recycling the symbol $I$, we denote the integrals within the sum as
\begin{equation}I_{2n-1}(\rho)=\int_\rho^1 \sqrt{\eta^2-\rho^2}\eta^{2n-1}d\eta\label{eq:DI}\end{equation}
for $n=0,1,2...$

\

For $n=0$, we may evaluate $I_{-1}$ using the substitution $\rho/\eta=\cos\theta$ and integration by parts
\begin{align}
I_{-1}(\rho)=\int_\rho^1 \frac{\sqrt{\eta^2-\rho^2}}{\eta}d\eta&=\rho \int_0^{\text{acos}(\rho)} \frac{\sin^2\theta}
{\cos^2\theta}d\theta \notag \\[9 pt]
&=\rho\left[\left(\frac{\sin\theta}{\cos\theta}\right | _0^{\text{acos}(\rho)}-\int_0^{\text{acos}(\rho)}d\theta\right]\notag \\[9 pt]
&=\sqrt{1-\rho^2}-\rho\,\text{acos}(\rho)\label{eq:DIm1}
\end{align}

The remaining integrals in the sum ($n>0$) can be found by recursion. Letting $m=2n-1$ and performing integration by parts
\begin{align*}
I_{m}(\rho)&=\int_\rho^1 \sqrt{\eta^2-\rho^2}\eta^{m}d\eta\\[9 pt]
&=\frac{1}{3}\left.(\eta^2-\rho^2)^{3/2}\eta^{m-1}\right|_\rho^1-\frac{m-1}{3}\int_\rho^1(\eta^2-\rho^2)^{3/2}\eta^{m-2}d\eta\\[9 pt]
&=\frac{1}{3}(1-\rho^2)^{3/2}-\frac{m-1}{3}\int_\rho^1 (\eta^2-\rho^2)^{1/2}\eta^m(\eta^2-\rho^2)\eta^{-2}d\eta\\[9 pt]
&=\frac{1}{3}(1-\rho^2)^{3/2}-\frac{m-1}{3}\left(I_m(\rho)-\rho^2 I_{m-2}(\rho)\right)
\end{align*}
which leads to the recursion relation
$$I_{m}(\rho)=\frac{(1-\rho^2)^{3/2}}{m+2}+\frac{m-1}{m+2}\rho^2 I_{m-2}(\rho)$$
with $I_1(\rho)=(1-\rho^2)^{3/2}/3$. We may solve this recurrence relation to yield an explicit expression for $I_m(\rho)$. Given the recursion relation, we can expect a solution of the form
$$I_m(\rho)=\frac{(1-\rho^2)^{3/2}}{m+2}\sum_{k=1,3,5...}^m a_{km}\rho^{k-1}$$
where
\begin{align*}
a_{1m}&=1\\
a_{3m}&=\frac{(m-1)}{m}\\
a_{5m}&=\frac{(m-1)}{m}\cdot\frac{(m-3)}{(m-2)}\\
a_{7m}&=\frac{(m-1)}{m}\cdot\frac{(m-3)}{(m-2)}\cdot \frac{(m-5)}{(m-4)}\\
...\\
a_{mm}&=\frac{(m-1)}{m}\cdot\frac{(m-3)}{(m-2)}\cdot \frac{(m-5)}{(m-4)}\cdot...\cdot\frac{2}{3}\\
\end{align*}
For convenience, we pass from odd-numbered $m$ and $k$ to a regular series of integers via $m=2n+1$ and $k=2j+1$ with $n$ and $j$ taking on the values $0,1,2,3,...$ such that we may write the integral $I_{m}$ as
$$I_{2n+1}(\rho)=\frac{(1-\rho^2)^{3/2}}{2n+3}\sum_{j=0}^n c_{jn}\rho^{2j}$$
with
\begin{align*}
c_{0n}&=1\\
c_{1n}&=\frac{(2n)}{(2n+1)}\\
c_{2n}&=\frac{(2n)}{(2n+1)}\cdot\frac{(2n-2)}{(2n-1)}\\
c_{3n}&=\frac{(2n)}{(2n+1)}\cdot\frac{(2n-2)}{(2n-1)}\cdot \frac{(2n-4)}{(2n-3)}\\
...\\
c_{nn}&=\frac{(2n)}{(2n+1)}\cdot\frac{(2n-2)}{(2n-1)}\cdot \frac{(2n-4)}{(2n-3)}\cdot...\cdot\frac{2}{3}
\end{align*}
Given experience with similar products in preceding appendices, we may express this as 
\begin{equation}
c_{jn}=\frac{2^n n!}{2^{n-j}(n-j)!} \cdot \frac{(2(n-j)+1)!/(2^{n-j}(n-j)!)}{(2n+1)!/(2^n n!)}
\label{eq:cj}
\end{equation}
where the term before $\cdot$ provides the products found in the numerator of the preceding expressions for $c_j$ whereas the term following $\cdot$ provides that found in the denominator. Proceeding to further simplify this expression for $c_j$
\begin{align*}
c_{jn}&=\frac{2^{2n} (n!)^2}{2^{2(n-j)}((n-j)!)^2} \cdot \frac{(2(n-j)+1)!}{(2n+1)!}\\[9pt]
&=\frac{\Gamma(n+1)}{\Gamma(n+3/2)}\frac{\Gamma(n-j+3/2)}{\Gamma(n-j+1)}
\end{align*}
where the first line simply rearranges (\ref{eq:cj}) and the latter makes use of the Gamma function and its the Legendre duplication formula (\ref{eq:Leg}). We thus arrive at the sought-after expression for $I_{2n+1}$, for $n=0,1,2...$. However, to be consistent with our use of $n$ in our original definition of the integral $I$, (\ref{eq:DI}) we shift $n$ and $j$ by one, such that $n$ and $j$ now take on values $1,2,3,...$, and 
$$I_{2n-1}(\rho)=\frac{(1-\rho^2)^{3/2}}{2n+3} \frac{\Gamma(n)}{\Gamma(n+1/2)}\sum_{j=1}^n \frac{\Gamma(n-j+3/2)}{\Gamma(n-j+1)}\rho^{2j-2}$$
Such that, upon substitution of this expression and that for $I_{-1}(\rho)$, \ref{eq:DIm1}, into our the expression for slip in the marginally pressurized limit, $\lambda\rightarrow 0$, 
\begin{equation*}
\frac{\delta(\rho,t)}{R(t)f \Delta p/\mu}=\frac{8}{\pi}\int_{\rho}^1 \frac{\sqrt{\eta^2-\rho^2}}{\eta}d\eta+\frac{4}{\sqrt{\pi}}\sum_{n=1}^\infty\lambda^{2n}\frac{(-1)^n}{\Gamma(n+3/2)}\int_{\rho}^1 \sqrt{\eta^2-\rho^2}\eta^{2n-1}d\eta
\end{equation*}
leads to
\begin{align*}
\frac{\delta(\rho,t)}{R(t)f \Delta p/\mu}=&\frac{8}{\pi}\left(\sqrt{1-\rho^2}-\rho\text{ acos}(\rho)\right)\\[9 pt]
&+\frac{4}{\sqrt{\pi}}\left(1-\rho^2\right)^{3/2}\sum_{n=1}^\infty \frac{\lambda^{2n}}{2n+1}\frac{(-1)^n}{\Gamma(n+3/2)}\frac{\Gamma(n)}{\Gamma(n+1/2)}\sum_{j=1}^n \frac{\Gamma(n-j+3/2)}{\Gamma(n-j+1)}\rho^{2j-2}
\end{align*}

\section{Far-field expansion of inner solution}
\label{app:d0}
In the interest of determining the overlap of the inner and outer solutions, here we derive the asymptotic expansion of the inner solution (\ref{eq:dinf}--\ref{eq:d0}) in the far field ($s\rightarrow\infty$). Specifically, we focus on the leading-order solution (\ref{eq:d0}), reproduced below,
\begin{equation*}\delta_0(s,t)=2\sqrt{\pi}e^{-s^2/2} \left[(1+s^2)I_0(s^2/2)+s^2 I_1(s^2/2)\right]-4s\end{equation*}
which follows from its integral representation (\ref{eq:d0OG})
\begin{equation}\delta_0(s,t)=\frac{8}{\pi}\int_s^\infty \sqrt{1-(s/\xi)^2}F(\xi)d\xi\label{eq:d0int}\end{equation}
where $F(\xi)$ is given by (\ref{eq:Fxi}).

\

We begin by substituting the asymptotic expansion of $F(\xi)$ for large values of its argument, (\ref{eq:F3}),
\begin{equation}F(\xi)\sim\sum_{n=0}^\infty \frac{1}{\sqrt{\pi}}\frac{1}{\xi^{2n+2}}\frac{\Gamma(n+3/2)}{2n+1}\label{eq:Fxi1}\end{equation}
into (\ref{eq:d0int}), which then requires us to evaluate integrals of the form 
$$\int_s^\infty \frac{\sqrt{1-(s/\xi)^2}}{\xi^{2n+2}}d\xi$$
which suggests the change of variable $\cos\theta=s/\xi$, transforming this integral to
$$\frac{1}{s^{2n+1}}\int_0^{\pi/2}\sin^2\theta\cos^{2n}\theta \,d\theta$$
the integral in which is a special case of one encountered in Appendix \ref{app:out} 
$$I(m,p,\rho)=\int_0^{\text{acos}(\rho)}\cos^m\theta\sin^p\theta \, d\theta$$
The particular solution for $p=2$ was found there to be 
$$I(m,2,\rho)= \frac{1}{2\sqrt{\pi}}\frac{\Gamma\!\left(\frac{m}{2}+\frac{1}{2}\right)}{\Gamma\!\left(\frac{m}{2}+2\right )}\left[\left(\text{acos}(\rho)-\rho\sqrt{1-\rho^2}\right)+\frac{(1-\rho^2)^{3/2}}{\rho}\sum_{k=2,4,...}^m \frac{2\sqrt{\pi}}{k+2}\frac{\Gamma\!\left(\frac{k}{2}+2\right)}{\Gamma\!\left(\frac{k}{2}+\frac{1}{2}\right)}\rho^k\right]$$
which simplifies for $\rho=0$ and $m=2n$ to 
$$I(2n,2,0)=\frac{\sqrt{\pi}}{4}\frac{\Gamma(n+1/2)}{\Gamma(n+2)}$$
Thus, following the substitution of (\ref{eq:Fxi1}) into (\ref{eq:d0int}), we deduce the asymptotic expansion for $\delta_0$ as $s\rightarrow \infty$
\begin{align}
\begin{split}
\delta_0(s,t)&\sim\frac{8}{\pi}\sum_{n=0}^\infty \frac{1}{\sqrt{\pi}}\frac{\Gamma(n+3/2)}{2n+1}\int_s^\infty\frac{\sqrt{1-(s/\xi)^2}}{\xi^{2n+2}}d\xi \\[9 pt]
&\sim\frac{2}{\pi}\sum_{n=0}^\infty \frac{\Gamma(n+3/2)}{2n+1}\frac{\Gamma(n+1/2)}{\Gamma(n+2)}\frac{1}{s^{2n+1}}
\end{split}
\end{align}

\medskip

\noindent{\bf References}
\small 
\medskip

\noindent Atkinson, C., and R. V. Craster (1994) A singular perturbation approach to integral equations occurring in poroelasticity, \emph{IMA J. Appl. Math.}, 52, 221--252, doi:10.1093/imamat/52.3.221

\medskip

\noindent Bhattacharya, P., and R. C. Viesca (2019) Fluid-induced aseismic fault slip outpaces pore-fluid migration, \emph{Science}, 364, 464--468, doi:10.1126/science.aaw7354

\medskip

\noindent Bender, C. M. and S. A. Orszag (1999) Advanced Mathematical Methods for Scientists and Engineers, \emph{Springer-Verlag}, doi:10.1007/978-1-4757-3069-2

\medskip

\noindent Berrios-Rivera, N., S. Ozawa, E. M. Dunham (2025) Models of fluid-induced seismic slip with permeability enhancement and rate-and-state friction, \emph{ESS Open Archive}, July 26, 2025, doi:10.22541/essoar.175357075.53109087/v1

\medskip

\noindent Craster, R. V., and C. Atkinson (1992) Shear cracks in thermoplastic and poroelastic media, \emph{J. Mech. Phys. Solids}, 40(4), 887--924, doi:10.1016/0022-5096(92)90008-P

\medskip

\noindent  Craster, R. V., and C. Atkinson (1994) Crack problems in a poroelastic medium: an asymptotic approach, \emph{Proc. Roy. Soc. Lond. A} 346, 387--428, doi:10.1098/rsta.1994.0026. 

\medskip

\noindent Desroches, J., E. Detournay, B. Lenoach, P. Papanastasiou, J. R. A. Pearson, M. Thiercelin, and A. Cheng (1994) The crack tip region in hydraulic fracturing, \emph{Proc. R. Soc. Lond. A}, 447, 39--48, doi:10.1098/rspa.1994.0127

\medskip

\noindent Dublanchet, P. (2019) Fluid driven shear cracks on a strengthening rate-and-state frictional fault, \emph{J. Mech. Phys. Solids}, 132, 103672, doi: 10.1016/j.jmps.2019.07.015

\medskip

\noindent Dunham, E. M. (2024) Fluid-driven aseismic fault slip with permeability enhancement and dilatancy, \emph{Phil. Trans. Roy. Soc. A} 382, 20230255, doi: 10.1098/rsta.2023.0255.

\medskip

\noindent Dyskin, A. V., L. N. Germanovich, and K. B. Ustinov (2000) Asymptotic analysis of crack interaction with free boundary, \emph{Int. J. Solids Struct.}, 37, 857--886, doi:10.1016/S0020-7683(99)00063-3

\medskip

\noindent Edmunds, T. M., and J.R. Willis (1976) Matched asymptotic expansions in nonlinear fracture mechanics—I. longitudinal shear of an elastic perfectly-plastic specimen, \emph{J. Mech. Phys. Solids}, 24(4), 205--223, doi:10.1016/0022-5096(76)90003-X

\medskip

\noindent Edmunds, T. M., and J.R. Willis (1976) Matched asymptotic expansions in nonlinear fracture mechanics—II. Longitudinal shear of an elastic work-hardening plastic specimen,\emph{J. Mech. Phys. Solids}, 24(4), 225--237, doi:10.1016/0022-5096(76)90004-1

\medskip

\noindent Edmunds, T. M., and J.R. Willis (1977) Matched asymptotic expansions in nonlinear fracture mechanics—III. In-plane loading of an elastic perfectly-plastic symmetric specimen, \emph{J. Mech. Phys. Solids}, 25(6), 423--455, doi:10.1016/0022-5096(77)90028-X

\medskip

\noindent Erickson, B. A., J. Jiang, M. Barall, N. Lapusta, E. M. Dunham, R. Harris, L. S. Abrahams, K. L. Allison, J.-P. Ampuero, S. Barbot, C. Cattania, A. Elbanna, Y. Fialko, B. Idini, J. E. Kozdon, V. Lambert, Y. Liu, Y. Luo, X. Ma, M. B. McKay, P. Segall, P. Shi, M. van den Ende, and M. Wei (2020) The community code verification exercise for simulating sequences of earthquakes and aseismic slip (SEAS), \emph{Seismol. Res. Lett.} 91(2A), 874--890, doi:10.1785/0220190248

\medskip

\noindent Garagash, D. I., and E. Detournay (2000) The tip region of a fluid-driven fracture in an elastic medium, \emph{J. Appl. Mech.}, 67, 183--192, doi:10.1115/1.321162.

\medskip

\noindent Garagash, D. I., and E. Detournay (2005) Plane-strain propagation of a fluid-driven fracture: small toughness solution, \emph{J. Appl. Mech.},  72(6), 916-928, doi:10.1115/1.2047596

\medskip

\noindent Garagash, D. I. (2006) Propagation of a plane-strain hydraulic fracture with a fluid lag: early-time solution, \emph{Int. J. Solids Struct.}, 43, 5811--5835, doi:10.1016/j.ijsolstr.2005.10.009

\medskip

\noindent Garagash, D. I., and L. N. Germanovich (2012) Nucleation and arrest of dynamic slip on a pressurized fault, \emph{J. Geophys. Res.},  117, B10310, doi:10.1029/2012JB009209

\medskip

\noindent Harris, R. A. , M. Barall, R. Archuleta, E. Dunham, B. Aagaard, J. P. Ampuero, H. Bhat, V. Cruz-Atienza, L. Dalguer, P. Dawson, S. Day, B. Duan, G. Ely, Y. Kaneko, Y. Kase, N. Lapusta, Y. Liu, S. Ma, D. Oglesby, K. Olsen, A. Pitarka, S. Song, E. Templeton (2009) The SCEC/USGS dynamic earthquake rupture code verification exercise, \emph{Seismol. Res. Lett.}, 80(1): 119--126, doi: 10.1785/gssrl.80.1.119

\medskip

\noindent Hinch, E. J. (1991), Perturbation Methods, \emph{Cambridge University Press}, doi:10.1017/CBO9781139172189

\medskip

\noindent Im, K. and J.-P. Avouac (2024), Maximum magnitude of induced earthquakes in rate and state friction framework, \emph{Seismol. Res. Lett.}, 6(3), 1654–1664, doi:10.1785/0220240382.

\medskip

\noindent Jacquey, A. B. and R. C. Viesca (2023) Nucleation and arrest of fluid-induced aseismic slip, \emph{Geophys. Res. Lett.}, 50, e2022GL101228, doi:10.1029/2022GL101228

\medskip

\noindent Lambert, V. R.,  B. A. Erickson, J. Jiang, E. M. Dunham, T. Kim, J.‐P. Ampuero, R. Ando, F. Cappa, P. Dublanchet, A. Elbanna, Y. Fialko, A.‐A. Gabriel, N. Lapusta, M. Li, J. Marcum, D. May, M. S. Mia, S. Ozawa, C. Pranger, P. Romanet, M. M. Scuderi, Y. van Dinther, Y. Yang, J. Yun (2025) Community‐driven code comparisons for simulations of fluid‐induced aseismic slip, \emph{J. Geophys. Res.}, 130, e2024JB030601, doi:10.1029/2024JB030601

\medskip

\noindent Lister, J. R.(1990) Buoyancy-driven fluid fracture: similarity solutions for the horizontal and vertical propagation of fluid-filled cracks, \emph{J. Fluid Mech.}, 217, 213--239, doi:10.1017/S0022112090000696

\medskip

\noindent Lister, J. R., D. J. Skinner, and T. M. J. Large, (2019) Viscous control of shallow elastic fracture: peeling without precursors, \emph{J. Fluid. Mech.}, 868, 119--140, doi:10.1017/jfm.2019.185

\medskip

\noindent Marck, J., A. A. Savitski, and E. Detournay (2015), Line source in a poroelasic layer bounded by an elastic space, \emph{Int. J. Numer. Anal. Meth. Geomech}, 39, 1484--1505, doi:10.1002/nag.2405

\medskip

\noindent Rice, J. R. (1980) Mechanics of earthquake rupture, in \emph{Physics of the Earth's Interior}, eds. A. M. Dziewonski and E. Boschi, Italian Physical Society/North Holland Publ. Co., pp. 555--649.

\medskip

\noindent Ross, Z. E., E. S. Cochran, D. T. Trugman, and J. D. Smith (2020) 3D fault architecture controls the dynamism of earthquake swarms, \emph{Science} 368, 1357--1361, doi.org/10.1016/0022-5096(91)90003-7

\medskip

\noindent Rudnicki, J. W. (1990) Boundary layer analysis of plane strain shear cracks propagating steadily on an impermeable plane in an elastic diffusive solid, \emph{J. Mech. Phys. Solids} 39(2), 201--221, doi.org/10.1016/0022-5096(91)90003-7

\medskip

\noindent Rudnicki, J. W., and D. A. Koutsibelas (1991) Steady propagation of plane strain shear cracks on an impermeable plane in an elastic diffusive solid, \emph{Int. J. Solids Struct.}, 27(2), 205--225., doi:10.1016/0020-7683(91)90229-9

\medskip

\noindent  S\'aez, A., B. Lecampion, P. Bhattacharya, and R. C. Viesca (2022) Three-dimensional fluid-driven stable frictional ruptures, \emph{J. Mech. Phys. Solids.}, 160, 104754, doi:10.1016/j.jmps.2021.104754

\medskip

\noindent  S\'aez, A., and B. Lecampion (2023) Post-injection aseismic slip as a mechanism for the delayed triggering of seismicity, \emph{Proc. Roy. Soc. A}, 479, doi:10.1098/rspa.2022.0810

\medskip

\noindent Sáez, A., and Lecampion, B. (2024) Fluid-driven slow slip and earthquake nucleation on a slip-weakening circular fault. \emph{J. Mech. Phys. Solids}, 183, 105506, doi:10.1016/j.jmps.2023.105506

\medskip

\noindent S\'aez, A., F. Passel\`egue, and B. Lecampion (2025) Maximum size and magnitude of injection-induced slow slip events, \emph{Sci. Adv.}, 11, eadq0662, doi: 10.1126/sciadv.adq0662

\medskip

\noindent  Salamon, N. J., and Dundurs, J. (1971) Elastic fields of a dislocation loop in a two-phase material, \emph{J. Elasticity}, 1(2), 153--164, doi:10.1007/BF00046466

\medskip

\noindent  Salamon N. J., and Dundurs, J. (1977) A circular glide dislocation loop in a two-phase material, \emph{J. Phys. C Solid State}, 10, 497--507, doi:10.1088/0022-3719/10/4/007

\medskip

\noindent Savitski, A. A., and E. Detournay (2002) Propagation of a penny-shaped fluid-driven fracture in an impermeable rock: asymptotic solutions, \emph{Int. J. Solids Struct.}, 39, 6311--6337, doi:10.1016/S0020-7683(02)00492-4

\medskip

\noindent Simons, D. A. (1977) Boundary-layer analysis of propagating mode II cracks in porous elastic media. \emph{J. Mech. Phys. Solids}, 25, 99-115, doi:10.1016/0022-5096(77)90006-0

\medskip

\noindent Simons, D. A. and J. R. Rice (1976) The stabilization of spreading shear faults by coupled deformation-diffusion effects in fluid-infiltrated porous materials, \emph{J. Geophys. Res.} , 81(29) 5322--5334.

\medskip

\noindent Sneddon, I. N. (1951) Fourier Transforms, \emph{McGraw-Hill}.

\medskip

\noindent Spence, D. A., and P. Sharp (1985) Self-similar solutions for elastohydrodynamic cavity flow, \emph{Proc. R. Soc. Lond. A}, 400, 289--313, doi:10.1098/rspa.1985.0081

\medskip

\noindent Thouless, M. D., A. G. Evans, M. F. Ashby, and J. W. Hutchinson (1987) The edge cracking and spalling of brittle plates, \emph{Acta Metall.} 35(6), 1333--1341, doi:10.1016/0001-6160(87)90015-0

\medskip

\noindent Viesca, R. C., and J. R. Rice (2012) Nucleation of slip-weakening rupture instability in landslides
by localized increase of pore pressure, \emph{J. Geophys. Res.}, 117, B03104, doi:10.1029/2011JB008866

\medskip

\noindent Viesca, R. C. (2021) Self-similar fault slip in response to fluid injection, \emph{J. Fluid Mech.}, 928, A29, doi:10.1017/jfm.2021.825

\medskip

\noindent Zlatin, A. N., and A. A. Khrapkov (1986) A semi-infinite crack parallel to the boundary of the elastic half-plane, \emph{Sov. Dokl. Phys.}, 31(12), 1009--1010.

\end{document}